\documentclass[twocolumn,showpacs,preprintnumbers,amsmath,amssymb]{revtex4}

\usepackage{graphicx}
\usepackage{revsymb}
\usepackage{dcolumn}
\usepackage{bm}

\begin{document}
\title{Laser control for the optimal evolution \\ of pure quantum states}
\author{D. Sugny$^1$}
\email{dominique.sugny@u-bourgogne.fr}
\author{A. Keller$^2$}
\author{O. Atabek$^2$}
\author{D. Daems$^3$}
\author{C. M. Dion$^4$}
\author{S. Gu\'{e}rin$^1$}
\author{H. R. Jauslin$^1$}
\affiliation{$^1$ Laboratoire de Physique de l'Universit\'{e} de Bourgogne, UMR CNRS 5027, BP 47870, 21078 Dijon, France}
\affiliation{$^2$ Laboratoire de Photophysique Mol\'{e}culaire du CNRS, Universit\'{e} Paris-Sud, B\^{a}t. 210 - Campus d'Orsay, 91405 Orsay Cedex, France}
\affiliation{$^3$ Center for Nonlinear Phenomena and Complex Systems, Universit\'{e} Libre de Bruxelles, 1050 Brussels, Belgium}
\affiliation{$^4$ Department of Physics, Ume\aa\ University, SE-90187 Ume\aa, Sweden}

\begin{abstract}
  Starting from an initial pure quantum state, we present a strategy
  for reaching a target state corresponding to the extremum (maximum
  or minimum) of a given observable. We show that a sequence of pulses
  of moderate intensity, applied at times when the average of the
  observable reaches its local or global extremum, constitutes a
  strategy transferable to different control issues. Among them,
  post-pulse molecular alignment and orientation are presented as
   examples. The robustness of such strategies with
  respect to experimentally relevant parameters is also examined.
\end{abstract}
\pacs{33.80.-b, 32.80.Lg, 42.50.Hz}
\maketitle

\section{Introduction}
Recent advances in laser technology have opened up large possibilities
for the control of quantum processes playing a role in a variety of
problems encompassing chemical reactivity \cite{warrer,seideman},
high-order harmonic generation \cite{atabek}, or logical gates with
applications in quantum information
\cite{nielsen,cirac,fonseca}. Several advances have been achieved in
this context, from the discovery of elementary basic
mechanisms of strong field induced molecular dynamics, to
optimal control strategies bringing together sophisticated numerical
algorithms (such as evolutionary algorithms)
\cite{atabek,dion}. Parallel to these works, a systematic study of
quantum systems in either pure or mixed (statistical) states has been
undertaken using control theory \cite{rabitz}. Upper and lower bounds
corresponding to kinematical constraints have been established for the
expectation value of arbitrary observables of driven quantum systems
\cite{girardeau,girardeau1}, with a particular attention paid to their
dynamical realizability \cite{rama,fu}. This latter question, which is
related to their complete controllability, has been solved for linear
control, i.e., for Hamiltonians depending in a linear way on the
control functions, of non-dissipative finite-level quantum systems
\cite{rama}.

The aim of this study, which is a continuation of our previous paper
\cite{sugny}, is to show how a strategy worked out for the laser
control of a specific observable can be transposed to a generic
system. An essential feature of the control scheme we propose is to
reduce the physical Hilbert space to a subspace $\mathcal{H}^{(N)}$,
of finite dimension $N$. A target state $|\chi ^{(N)}\rangle $, which
maximizes (or minimizes) the expectation value of an observable
$\mathcal{O}$, is then defined in this subspace. The first obvious
advantage of the finite dimensionality being that $|\chi ^{(N)}\rangle
$ appears, through a variational principle, to be the eigenvector (for
the non-degenerate case) associated to $\mathcal{H}^{(N)}$ with the
highest or lowest eigenvalue of the restriction of the observable
$\mathcal{O}$. These bounds being kinematic, they only indicate to
which extent the system can be dynamically controlled by application
of time-dependent external fields. Moreover, in the finite-dimensional
case, the complete or non-complete controllability can be determined
and related (for a linear control system) to the dimension of the Lie
algebra generated by the unperturbed Hamiltonian and the interaction
operator \cite{rama,fu}.

Assuming the dynamical realizability of the target state, the
remaining question consists in explicitly finding a suitable scheme
in terms of a unitary evolution operator $U(t)$ which, starting from
an initial pure state of the molecular system, dynamically brings it
close to the state $|\chi ^{(N)}\rangle $. Several control strategies
have already been proposed using, for instance, adiabatic passage
techniques \cite{vitanov,jauslin}, factorizations of unitary operators
\cite{schirmer} or optimal control schemes \cite{dion,zhu}. Here, we
pursue another approach and we show that a sequence of short pulses of
moderate intensity can be devised for reaching these target states. A
generalization to models incorporating fluctuating environmental
effects such as temperature, via statistically mixed initial states,
is still an open question on which work is in progress.

The strategy is exemplified by two illustrative cases, the alignment
and the orientation control of rigid-rotor molecules. We show that
either complete or close to complete control is achieved after the
application of a finite number of pulses. More precisely, the
alignment / orientation control processes, aiming both at efficiency
and maximal duration within a rotational period, require few kicks to
get close to the target, while remaining in the conventionally fixed
finite dimensional Hilbert subspace. We note that in all the dynamical
mechanisms under consideration, we are interested in controlling the
post-pulse behavior of the observables. The motivations for such
processes are now well established in a wide variety of applications
extending from chemical reaction dynamics to surface processing,
catalysis, nanoscale design, and quantum computing
\cite{seideman,brooks,seideman1,seideman2,aoiz}.

The paper is organized as follows. A mathematical setup is given in
Sec.~\ref{section2} for the general control strategy, which is then
applied to the case of molecular alignment / orientation. These
processes are taken on a parallel footing in Sec.~\ref{section4} by
considering either the polarizability or the permanent dipole moment
of the molecular system. Section \ref{section5} collects the results
with a special emphasis on the robustness with respect to control
parameters and the experimental feasibility. Concluding remarks and
prospective views are presented in Sec.~\ref{section6}. Some details
of the calculations and some proofs are presented in Appendices
\ref{appa}, \ref{appb}, \ref{appc}, and \ref{appd}.

\section{Methodology} \label{section2} 
The field-free Hamiltonian acting on the physical Hilbert space
$\mathcal{H}$ is denoted $H_0$. The initial pure quantum state of
interest $|\psi_n (t=0)\rangle$, is taken as one of the eigenstates
of $H_0$ corresponding to the eigenvalue $E_n$. One considers an
observable $\mathcal{O}$ which, through the dynamical behavior of its
expectation value $\langle \mathcal{O}(t)\rangle$, describes the
physical property to be controlled by the external
field. $\mathcal{O}$ is a self-adjoint operator which does not commute
with $H_0$. The aim of the control is to find an evolution operator
$U(t)$, within a class of experimentally realizable processes, such
that the time evolution of the initial state under the action of
$U(t)$ leads to an average $\langle \psi_n
(t)|\mathcal{O}|\psi_n(t)\rangle$ that is maximized or minimized. This
goal is achieved in three steps.

\subsection{Finite dimensional subspace} \label{section2A}

We first select a finite $N$-dimensional subspace $\mathcal{H}^{(N)}$
of the physical Hilbert space. For instance, in the case of alignment
/ orientation such a subspace is generated by the first $N$
eigenstates $|n\rangle$ of $H_0$ with eigenvalues $E_n$
\cite{sugny}. The mathematical advantage of this reduction is twofold
when considering the reduced operator 
\begin{equation} \label{eq1}
\mathcal{O}^{(N)}=P^{(N)}\mathcal{O}P^{(N)} ,
\end{equation}
$P^{(N)}$ being the projector on the subspace $\mathcal{H}^{(N)}$. The
first advantage, of kinematical nature \cite{girardeau,girardeau1}, is
related to the fact that $\mathcal{O}^{(N)}$ has now an upper and
lower bounded discrete spectrum, as opposed to the possibly continuous
or unbounded spectrum of $\mathcal{O}$. The second is that the
controllability of the system can be completely analyzed (see
Sec.~\ref{section2C} for details), whereas such results are limited
for the infinite-dimensional case \cite{lloyd}.

Finally, we assume that all the fundamental frequencies of $H_0$ in
$\mathcal{H}^{(N)}$ are commensurate, i.e., integer multiples of a
fundamental frequency $\omega$, which implies that the motion is
periodic in time, with period $T=2\pi/\omega$.

Apart from these mathematical considerations, which will be used in
Sec.~\ref{section2C}, the reduction of the Hilbert space may also lead
to other physical advantages in specific cases. Indeed, for post-pulse
alignment / orientation control, the finite dimensionality of the
molecular rotational space, involving low momenta, leads to longer
alignment / orientation dynamics as a result of the slower oscillation
of $\langle \mathcal{O}(t) \rangle$ after interaction with the
field. It is precisely this behavior that roughly constitutes the
basis of the compromise between efficiency and long duration of such
control issues. Reciprocally, the justification of the dimensionality
reduction is related to the fact that moderate perturbations,
i.e., a moderate number of applied pulses and amplitude, can only
transfer finite amounts of energy to the molecule, confining the system
to a finite-dimensional subspace \cite{sugny}.

\subsection{Controllability} \label{sec_control}
The control is exerted through the application of a time-dependent
external field described by the Hamiltonian
\begin{equation} \label{eq4}
H(t)=H_0+V(t) ,
\end{equation}
where $V(t)$ is the molecule-field interaction potential. The general
structure of $V(t)=V(\vec{\xi},\vec{\mathcal{E}}(t))$ involves the
field vector $\vec{\mathcal{E}}(t)$ (amplitude and polarization
direction) together with molecular vectorial characteristics
$\vec{\xi}$ that may, in some cases, be field-induced (an example of
such a behavior, through the polarizability interaction, is given in
Sec.~\ref{section4}). Moreover, in order to apply the results of
control theory, we also assume that the system is control-linear, which
means here that $V$ can be written in the form 
\begin{equation} \label{eq4a}
V(t)=-v(t)H_I ,
\end{equation} 
where $v$ is a real control function (that can be chosen at will) and
$H_I$ the interaction operator. If a system of the form (\ref{eq4a}) is
completely controllable \cite{rama,fu} then there exists
a function $v(t)$ such that the corresponding propagator $U(t)$ brings
the initial state to the target state. A simple example of a
non-controllable system is a decoupled one. In this case, the matrix
representation of $H$ is block-diagonal and the Hilbert space can be
decomposed as a direct sum of at least two orthogonal
subspaces. However, for non-decoupled systems, the question of
dynamical realizability is more difficult and an investigation of the
structure of the Lie algebra is needed.  A Lie algebra of matrices is
a subspace of the vector space (over the field of reals) of $N \times
N$ matrices (with complex entries) which is stable under the
commutation operation \cite{rama}.  Two cases of particular
relevance here are the Lie algebra of $N \times N$ skew-Hermitian
matrices denoted by $u(N)$ and whose dimension is $N^2$, and the Lie
algebra $su(N)$ of $N \times N$ zero-trace skew Hermitian matrices,
whose dimension is $N^2-1$. We recall \cite{rama,fu} that a necessary
and sufficient condition for the complete controllability of the
Hamiltonian system defined by $H$ [Eq.~(\ref{eq4})] is that the Lie
algebra generated by $iH_0$ and $iH_I$ be $u(N)$. In the rest of this
section, we will assume the dynamical realizability of the target
states.

\subsection{Target state} \label{section2B}
The state $|\chi^{(N)}\rangle$ which maximizes (or minimizes)
$\mathcal{O}^{(N)}$ in the subspace $\mathcal{H}^{(N)}$ is nothing but
the eigenstate of $\mathcal{O}^{(N)}$ corresponding to its upper (or
lower) eigenvalue. This provides a clear description of the target
state in terms of a linear combination of states $|n\rangle$ in direct
relation with the observable to control. If $N$ is large enough, the
upper (or lower) eigenvalue of $\mathcal{O}^{(N)}$ will be close to
the corresponding bound of $\mathcal{O}$ (if this operator is bounded)
and an efficient control could be achieved (assuming complete
controllability). Once again, it is worth noting that large $N$ allows
for higher values of $|\langle \mathcal{O}^{(N)}\rangle |$, but leads
to higher energy transfer from the external field to the molecular
system and consequently results into shorter durations as the
post-pulse average $\langle \mathcal{O}(t) \rangle$ is done over a
superposition involving many more eigenstates of $H_0$.

\subsection{Control strategy} \label{section2C}

We shall now present in detail two control schemes.  The first one,
which was first proposed in \cite{averbuch} and further developed in
\cite{sugny}, consists in applying pulses each time $\langle
\mathcal{O}^{(N)}\rangle$ reaches its maximum value within a period
$T$ of the field-free dynamics. An individual pulse perturbs the
molecule by an effective evolution operator $U_{\tilde{A}}$, where
$\tilde{A}$ is a real parameter related to the field amplitude:
Eq.~(\ref{eq8}) gives, for instance, the expression of $\tilde{A}$ in
the sudden approximation. The system remains in the subspace
$\mathcal{H}^{(N)}$, up to negligible corrections, if the
field amplitude $\tilde{A}$ is small enough. Between two successive
pulses, the molecule evolves following its field-free motion, governed
by $U_{H_0}(\Delta t)=\exp (-iH_0 \Delta t)$. An important hypothesis
of our strategy is that the system is to be perturbed according to a
unitary operator $U_{\tilde{A}}$, which commutes with
$\mathcal{O}^{(N)}$, such that its application does not alter $\langle
\mathcal{O}^{(N)}\rangle =\langle U_{\tilde{A}}^{-1}\mathcal{O}^{(N)}
U_{\tilde{A}} \rangle$. Furthermore, the optimal target state $|\chi
^{(N)}\rangle$ is an eigenfunction of both $\mathcal{O}^{(N)}$ and
$U_{\tilde{A}}$.

We now consider that pulses are applied at times $t_i$ when $\langle
\mathcal{O}^{(N)}(t)\rangle $ reaches its global maximum
$\mathcal{O}_i=\langle \mathcal{O}^{(N)}(t_i)\rangle$. The sequence
of $\mathcal{O}_i$'s is increasing (possibly constant) but bounded and
is therefore convergent. Due to the hypothesis on the fundamental
frequencies (Sec.~\ref{section2A}) and the fact that $\langle
\mathcal{O}^{(N)}\rangle =\langle U_{\tilde{A}}^{-1}\mathcal{O}^{(N)}
U_{\tilde{A}} \rangle$, $\langle \mathcal{O}^{(N)}(t)\rangle$ is a
periodic, continuous and differentiable (under free-evolution)
function. It is then easy to verify that the limit of the preceding
sequence is a fixed point $\mathcal{O}_i=\mathcal{O}_{i+1}$,
corresponding to a wave function $|\psi_f\rangle$ (or a family of wave
functions) such that $\langle\psi_f|\mathcal{O}^{(N)}|\psi_f\rangle
=\langle\psi_f|U_{\tilde{A}}^{-1}\mathcal{O}^{(N)}U_{\tilde{A}}|\psi_f\rangle=\mathcal{O}_i$
is a global maximum within a period $T$ for any value of the parameter
$\tilde{A}$.

The description of this control strategy raises several questions
which are not completely resolved. The first one is the set of limits
(corresponding to the fixed points) of the process. It can readily be
shown that this set contains some of the eigenvectors of the operator
$\mathcal{O}^{(N)}$, but we can look for the conditions on $H_0$,
$U_{\tilde{A}}$ and $\mathcal{O}^{(N)}$ such that these wave functions
be the unique fixed points of the strategy. Appendix \ref{appa}
provides a sufficient condition on these operators. The second
question deals with the determination of the domain of attraction of
each fixed point and, more precisely, the domain of the target
state. The domain of attraction of $|\psi_f\rangle$ can be defined as
the set of initial states $|\psi\rangle$ such that, for some choice of
the parameters $\tilde{A}_i$ (the pulses being labeled by $i$), the
process, starting from $|\psi\rangle$ converges to
$|\psi_f\rangle$. The last open question concerns the reduction to a
finite dimensional space of the dynamics. In this paper, the fact that
the dynamics resides within the subspace $\mathcal{H}^{(N)}$ is only
numerically justified (see Sec.~\ref{section4} for details). Work is
in progress on these open questions.
 
The second control scheme that can be used is a strategy similar in
its spirit to the first one. Up to now, we have searched for a
sequence of perturbations that maximize the average value of the
observable $\langle \mathcal{O}^{(N)}(t)\rangle$. Another relevant
optimization would be to get closer to the target state $|\chi
^{(N)}\rangle$. This can be done in practice by applying pulses each
time the modulus of the projection on the target state, $|\langle \chi
^{(N)} |\psi _n (t)\rangle|$, reaches a maximum. The relative merits
of the two control schemes will be discussed in Sec.~\ref{section5}.

To conclude, we mention an alternative that can favor the convergence
of the control scheme. The general strategy is based on global maxima
of the field-free dynamics, but local maxima can also be used. This
will be illustrated numerically for the control of molecular alignment
in Sec.~\ref{section5}, where it will be shown that a better
convergence is obtained with local maxima.

\subsection{The sudden approximation} \label{section2D} 
Depending on the duration $\tau$ of the radiative interaction $V(t)$
as compared to molecular characteristic times $T$, we may be in the
adiabatic ($\tau\gg T$) or in the sudden ($\tau\ll T$) limits. Long
laser pulses have been shown to yield very efficient adiabatic
molecular control \cite{vrakking,guerin,friedrich,sangouard}, but with
the caveat of it disappearing after the pulse, as the molecule returns
to its initial state when the turn off is adiabatic. As we are rather
concerned with the post-pulse behavior of $\langle \mathcal{O}(t)\rangle
$, we hereafter use short pulses, assuming the sudden
limit. Following the derivation of the lowest order impulsive
approximation given in Refs.~\cite{dion2,henriksen}, the
time-dependent Schr\"{o}dinger equation, written in the
interaction representation as
\begin{equation} \label{eq5}
i\hbar \frac{\partial}{\partial t} \tilde{\psi}_\lambda(t) = -
\left[e^{iH_0t/\hbar}V(t)e^{-iH_0t/\hbar} \right] \tilde{\psi}_\lambda(t) ,
\end{equation}
with $\tilde{\psi}_n(t)=e^{iH_0t/\hbar}\psi_n(t)$, is integrated on
the total pulse duration $\tau$ to yield 
\begin{equation} \label{eq6}
\tilde{\psi}_n(\tau) = \frac{i}{\hbar} \int_0^\tau
\left[e^{iH_0t/\hbar}V(t)e^{-iH_0t/\hbar} \right]\tilde{\psi}_n(t)dt
+\tilde{\psi}_n(t=0)  . 
\end{equation}
At the lowest order in $\tau/T$, one can substitute $e^{iH_0t/\hbar}$
with $\textbf{1}$ in Eq.~(\ref{eq6}) and,  using $V(t)=-v(t)H_I$
[Eq.~(\ref{eq4a})], the integral equation (\ref{eq6}) can be solved as
\begin{equation} \label{eq7}
\tilde{\psi}_n(\tau)=\exp (i\tilde{A}H_I)\tilde{\psi}_n (t=0)  ,
\end{equation}
with
\begin{equation} \label{eq8}
\tilde{A}= \frac{1}{\hbar} \int_0^\tau v(t)dt  .
\end{equation}
Finally, returning to the Schr\"{o}dinger representation, one obtains
\begin{equation} \label{eq9}
\psi_n(t\geq\tau)=U_{H_0}(t)U_{\tilde{A}}\psi_n(t=0)  ,
\end{equation}
where the evolution operators are given by 
\begin{equation} \label{eq10}
U_{H_0}(t)=\exp (-iH_0t/\hbar) 
\end{equation}
and
\begin{equation} \label{eq11}
U_{\tilde{A}}=\exp (i\tilde{A}H_I)  .
\end{equation}
We are now in a position to examine the conditions (choice of the
parameters $N$, $A$, and $t_i$'s) under which the alluded strategies
work in the control of alignment / orientation processes.

\section{Control of the molecular alignment / orientation} \label{section4} 

Laser-induced alignment and orientation are important issues in
controlling molecular dynamics \cite{seideman}. For a diatomic
molecule driven by a linearly polarized laser field, alignment means
an increased probability distribution along the polarization axis
whereas orientation requires in addition the same (or opposite)
direction as the polarization vector. In a variety of applications,
extending from chemical reaction dynamics to surface processing,
catalysis and nanoscale design \cite{seideman,brooks,aoiz,seideman1},
noticeable orientation that persists after the end of the pulse is of
special importance. This is why the identification of target states
which fulfill these two requirements (i.e., efficiency and persistence)
is the basis of the construction presented in Ref.~\cite{sugny}.

The time evolution of the molecule, hereafter described in a 3D rigid
rotor approximation, interacting with a linearly-polarized field is
governed by the time-dependent Schr\"{o}dinger
equation (TDSE), expressed in atomic units,
\begin{equation} \label{eq35}
i\frac{\partial}{\partial t} \psi(\theta,\phi ;t) = \left[
  BJ^2-\vec{\mu}(\vec{\mathcal{E}}) \cdot \vec{\mathcal{E}}(t) \right]
\psi(\theta,\phi ;t) ,
\end{equation}
where $J$ is the angular momentum operator, $B$ the rotational
constant, $\vec{\mu}(\vec{\mathcal{E}})$ the total dipole moment and
$\vec{\mathcal{E}}$ the electric field. $\theta$ denotes the polar
angle between the molecular axis and the polarization direction of the
applied field. The motion related to the azimuthal angle $\phi$ can be
separated due to cylindrical symmetry. The field-induced dipole is
expanded in powers of $\vec{\mathcal{E}}(t)$,
\begin{equation} \label{eq36}
\vec{\mu}=\vec{\mu}_0+\bar{\bar{\alpha}} \cdot \vec{\mathcal{E}}(t)+\ldots ,
\end{equation}
where the two first terms (to which our approximation is limited) are
respectively the permanent dipole $\vec{\mu}_0$ and the polarizability
tensor $\bar{\bar{\alpha}}$. We notice that only very intense lasers
may require the inclusion in Eq.~(\ref{eq36}) of higher order
terms. The electric field is written as
\begin{equation} \label{eq37}
\vec{\mathcal{E}}(t)=f(t)\cos(\omega t) \vec{\epsilon} ,
\end{equation}
$\vec{\epsilon}$ being the unit polarization vector, $\omega$ the
carrier wave frequency and $f(t)$ the laser pulse envelope. In a
high-frequency regime, which is often referred to in models dealing
with alignment \cite{seideman}, the permanent dipole interaction,
i.e., $\vec{\mu}_0 \cdot \vec{\mathcal{E}}(t)$, averages to zero due
to the term $\cos(\omega t)$, whereas the polarizability interaction
corresponding to $\bar{\bar{\alpha}} \cdot \vec{\mathcal{E}}^2(t)$,
involves a $\cos^2(\omega t)$ term which contributes an average
$1/2$. Finally, in a high-frequency approximation
\cite{friedrich,keller}, the TDSE [Eq. (\ref{eq35})] is written as 
\begin{equation} \label{eq38}
i\frac{\partial}{\partial t} \psi(\theta,\phi;t) = \left[
  BJ^2- \frac{1}{4} f^2(t) \left(
    \Delta\alpha\cos^2\theta+\alpha_\perp \right)
\right ]\psi(\theta,\phi 
;t) , 
\end{equation}
where $\Delta\alpha=\alpha_\parallel-\alpha_\perp$ is the difference
between the parallel $\alpha_\parallel$ and perpendicular
$\alpha_\perp$ components of the polarizability tensor. While
excellent alignment can be obtained through adiabatic transport on
so-called pendular states \cite{friedrich} resulting from field
dressing, only sudden pulses offer the possibility of alignment that
persists after the field is over \cite{seideman}. As opposed to
alignment, orientation requires spatial symmetry breaking, and
therefore cannot be obtained from a radiative interaction in
$\cos^2\theta$ such as the one depicted in Eq.~(\ref{eq38}) resulting
from polarizability. It has recently been shown that very short pulses
combining a frequency $\omega$ and its second harmonic $2\omega$ (in
resonance with a vibrational transition), excite, through the combined
effect of $\mu_0$ and $\alpha$, a mixture of even and odd rotational
levels and have the ability to produce post-pulse orientation
\cite{dion1}. But even more efficient orientation is obtained using
half-cycle pulses (HCPs), that, through their highly asymmetrical shape,
induce a very sudden momentum transfer to the molecule which orients
under such a kick after the field is off \cite{dion2,machholm}. It is
worth noting that both the $(\omega+2\omega)$ and the kick mechanisms
have received a confirmation from optimal control schemes
\cite{dion}. In a moderate field approximation where the
polarizability interaction is neglected as compared to the permanent
dipole one, the TDSE which governs the orientation process can be
written as 
\begin{equation} \label{eq39}
i\frac{\partial}{\partial t} \psi(\theta,\phi ;t) = \left[
  BJ^2-f(t)\mu_0\cos\theta \right]\psi(\theta,\phi ;t) .
\end{equation}
As the two processes (i.e., post-pulse alignment / orientation) require
short duration pulses, from now on, we assume a sudden approximation
based on the shortness of the pulse duration $\tau$ as compared to the
molecular rotational period $T_{rot}=\pi/B$. For relatively low $j$
(where $j$ labels the eigenstates of $J^2$), this amounts to the
definition of a dimensionless, small perturbative parameter
$\varepsilon=\tau B$. This definition, together with a rescaling of
time $s=t/\tau$ (such that $s\in [0,1]$ during the pulse) leads to a
TDSE that can be treated by time-dependent unitary perturbation theory
\cite{sugny2,daems},
\begin{equation} \label{eq40}
i\frac{\partial}{\partial s} \psi(\theta,\phi ;s) =
\left[H_0-V_{a;o}(s) \right]\psi(\theta,\phi ;s) ,
\end{equation}
with $H_0=\varepsilon J^2$ and $V_{a;o}(s)$ the radiative interaction
depending on the process to be described.  More precisely, we have for
alignment
\begin{equation} \label{eq41}
V_a(s)=E_a^2(s)\cos^2\theta+F_a^2(s) ,
\end{equation}
and for orientation 
\begin{equation} \label{eq42}
V_o(s)=E_o(s)\cos\theta .
\end{equation}
The time-dependent functions are defined as 
\begin{eqnarray} \label{eq43}
E_a^2(s)&=& \Delta\alpha\tau f^2(\tau s)/2 ,\\ 
F_a^2(s)&=& \alpha_\perp \tau f^2(\tau s)/2 , \\
E_0(s) &=& \mu_0 \tau f(\tau s) .
\end{eqnarray}

Returning now to the successive steps of our approach as detailed in
the mathematical setup (Sec.~\ref{section2}), we have to define an
initial pure quantum state and the observables adequately describing
both processes. The initial state of the molecule is taken as the
ground state of the rigid rotor,
\begin{equation} \label{eq44}
|\psi_{n=0}(t=0)\rangle=|j=0,m=0\rangle .
\end{equation} 
We remark that any other rotational state of the rotor, i.e., $|j\geq
m,m\rangle$ could not be taken, by observing that the projection
quantum number $m$ is invariant upon the application of linearly
polarized pulses. The two observables we consider are
$\mathcal{O}_a=\cos^2\theta$ for alignment and
$\mathcal{O}_o=\cos\theta$ for orientation. The goal of the field
driven molecular alignment (or orientation) is then to maximize (or to
minimize, depending the choice of the orientation) for the largest
time duration, the expectation values
\begin{equation} \label{eq45}
\langle \mathcal{O}_{a;o}(s)\rangle =\langle \psi(\theta,\phi ;s)|\mathcal{O}_{a;o}|\psi(\theta,\phi; s)\rangle ,
\end{equation} 
after the pulse is over.

The second step consists in the reduction of the dimensionality of the
infinite physical Hilbert space by considering a finite subspace
$\mathcal{H}_{m=0}^{(N)}$ generated by the first $N$ eigenvectors of
$J^2$, i.e., $|j,m=0\rangle$ ($j=0,1,\ldots,N-1$). The observables are
projected in this subspace according to
\begin{equation} \label{eq45a}
\mathcal{O}_{a;o}^{(N)}=P_{m=0}^{(N)}\mathcal{O}_{a;o}P_{m=0}^{(N)} ,
\end{equation}
with the projectors explicitly given by
\begin{equation} \label{eq46}
P_{m=0}^{(N)}=\sum_{j=0}^{N-1} |j,m=0\rangle\langle j,m=0| .
\end{equation}
It is clear from Eq.~(\ref{eq45a}) that $\mathcal{O}_{a;o}^{(N)}$ has,
as opposed to $\cos^2\theta$ or $\cos\theta$, a purely discrete
spectrum. The target states $|\chi_{a;o}^{(N)}\rangle$ which maximize
the alignment (or the orientation) in the subspace
$\mathcal{H}_{m=0}^{(N)}$ are simply the eigenstates of
$\mathcal{O}_{a;o}^{(N)}$ with the highest eigenvalue. Their
calculation in terms of the appropriate expansion coefficients
$c_j^{a;o}$ in the $|j,m=0\rangle$ basis,
\begin{equation} \label{eq47}
|\chi_{a;o}^{(N)}\rangle=\sum_{j=0}^{N-1}c_j^{a;o}|j,m=0\rangle ,
\end{equation}
involves the diagonalization of the $(N\times N)$ matrices of
$\mathcal{O}_{a;o}^{(N)}$ [Eq.~(\ref{eq45a})] written in the same
basis. For instance, in the simplest case of orientation, a
tridiagonal matrix of the operator $\mathcal{O}_o^{(N)}$ results (when
using the approximation $\langle j,m=0|\cos\theta|j\pm 1,m=0\rangle
=1/2$, valid for $j\gg m$) in the analytical expression
\begin{equation} \label{eq48}
|\chi_{o}^{(N)}\rangle\simeq \left(\frac{2}{N+1} \right)^{1/2}
\sum_{j=0}^{N-1}\sin \left(\pi\frac{j+1}{N+1} \right)|j,m=0\rangle , 
\end{equation}
with the corresponding maximal orientation found in this subspace
\begin{equation} \label{eq49}
\langle \chi_o ^{(N)}|\cos\theta|\chi_o^{(N)}\rangle \simeq \cos
\left( \frac{\pi}{N+1} \right) .
\end{equation}

Having determined the kinematical constraints, we can now analyze the
dynamical realizability of these target states. This kind of
Hamiltonian (involving a dipole interaction with the control field)
has already been investigated by S. G. Schirmer and co-workers
\cite{fu} and it was shown that the Lie algebra generated by
$iH_0^{(N)}$ and $i\mathcal{O}_{a;o}^{(N)}$ is $u(N)$ (Theorem 1 of
Ref.~\cite{fu}). The system is therefore completely controllable.

The next step consists in applying our control strategy. For doing so,
we determine the evolution operator for individual pulses. This is
done in the impulsive sudden limit \cite{dion2,henriksen}. The
interaction term of the alignment process $V_a(s)$ [Eq.~(\ref{eq41})]
involves the square of the electric field amplitude. The
time evolution can be expressed in terms of a free evolution combined
with
\begin{equation} \label{eq50}
U_a=\exp \left[iA_a\cos ^2\theta \right] ,
\end{equation}
where
\begin{equation} \label{eq51}
A_a=\int _0 ^1 E_a^2(s)ds
\end{equation} 
is the total pulse-square area. We note that we have not taken into
account the $\theta$-independent contribution $F_a^2(s)$ to $V_a(s)$
as it would only contribute a pure phase factor. For the case of
orientation, we get as the instantaneous evolution operator describing
the kick 
\begin{equation} \label{eq52}
U_o=\exp \left[ iA_0\cos\theta \right] ,
\end{equation}
with
\begin{equation} \label{eq53}
A_o=\int _0^1 E_o(s)ds 
\end{equation}
the total pulse area. Between pulses, the molecule evolves under the
effect of its field-free rotation, described by the evolution operator
\begin{equation} \label{eq54}
U_{H_0}=\exp \left[ -i\varepsilon J^2 s \right] .
\end{equation}
One can easily check the necessary commutation relations
\begin{equation} \label{eq55}
\left[ \mathcal{O}_{a;o}^{(N)},U_{a;o} \right]=0 ,
\end{equation}
implying that the target states $| \chi_{a;o}^{(N)}\rangle$ are
eigenfunctions of both $\mathcal{O}_{a;o}^{(N)}$ and $U_{a;o}$. Note
that all quantities ($s$, $\varepsilon$, $E_a^2$, $E_o$) are
dimensionless, and hence do not depend on a specific molecular
system. Times can be expressed in terms of fractions of the molecular
rotational period $T_{rot}$. Moreover, the molecule-laser interaction
characteristics ($E_a^2$ or $E_o$) combine the field strength
$\mathcal{E}$ with molecular constants $\alpha$ or $\mu_0$, in such a
way that the same values may be considered for different molecules at
other, adequately chosen laser intensities. The final step for the
control scheme consists in applying the instantaneous perturbations at
specific times. More precisely, as has been discussed in the
mathematical setup (Sec.~\ref{section2}), two distinct strategies are
adopted for the determination of these times. The first one ($S1$)
consists in taking the series of $s_i$ when $\langle
\mathcal{O}_{a;o}^{(N)}\rangle (s)$ reaches its successive maxima
(global or local) during the free evolution following a kick. The
second one ($S2$) consists in taking the $s_i$'s as times when the
modulus of the projection of the wave function on the target states
$|\langle \psi_\lambda (s)|\chi_{a;o}^{(N)}\rangle|$ reaches a maximum
during the free evolution. The numerical results obtained by both
strategies are presented in Sec.~\ref{section5}.

We now discuss the relative merits of the two schemes from a
theoretical point of view.  For the strategy $S1$, the slope after the
pulse can be explicitly determined in the infinite-dimensional case,
as shown in Appendix \ref{appc},
\begin{equation} \label{eq550}
\left. \frac{d}{ds}\langle \cos \theta \rangle \right|_{s_i+0} =2\varepsilon A
(\textbf{1}-\langle\cos^2 \theta\rangle ) , 
\end{equation}
for orientation, and 
\begin{equation} \label{eq551}
\left. \frac{d}{ds}\langle \cos^2 \theta \rangle \right|_{s_i+0}
=2A\varepsilon\langle \sin ^2 2\theta\rangle  ,
\end{equation}
for alignment. As $\langle \cos ^2 \theta\rangle <1$ and $\langle \sin
^2 2\theta\rangle >0$, one deduces that, in both cases, the slope
takes a non-zero value after the application of the kick, whatever the
pulse area $A$ of the field.  This point proves that the control
scheme can be iterated for the alignment / orientation processes. In
the finite dimensional case, the corresponding expression for
orientation derived in Appendix \ref{appc} is slightly different. We
obtain for $\cos^{(N)}\theta$ 
\begin{equation} \label{eq552}
\left. \frac{d}{ds}\langle \cos^{(N)} \theta \rangle \right|_{s_i+0}
=2\varepsilon A (\textbf{1}-\langle(\cos^{(N)}
\theta)^2\rangle+B)+O(A^2) , 
\end{equation}
$B$ being a boundary term which is of the form 
\begin{equation}
B=-\frac{N^2+2N+1}{2N+1} |a_N|^2 ,
\end{equation}
where the $a_n$'s are the coefficients of the wave function in the
basis $|0\leq j\leq N-1,m=0\rangle$. Here, we notice that we cannot
conclude as above and arguments of dimensionality have to be used. In
this way, Appendices \ref{appa} and \ref{appb} show that the unique
fixed point of strategy $S2$ is the target state $|\chi^{(N)}\rangle$
whereas scheme $S1$ can possess several fixed points as the
eigenfunctions of the operator $\mathcal{O}^{(N)}$. However, the
strategy $S1$ has the advantage that the target state can be selected
during the dynamical process by appropriately choosing the intensities
and the number of laser pulses. This is not the case for the scheme
$S2$ because the target state has to be selected a priori.

\section{Results} \label{section5} 
The results are presented in a general way, transferable to any
particular molecule (characterized by its rotational period, its
permanent dipole and its polarizability) interacting with any pulsed
field (characterized by its moderate amplitude and its short duration
as compared to the molecular rotational period). The relevant
parameters which gather both molecule and field constants are taken to
be $A_{a}=1.5$, $A_{o}=1$, and $\varepsilon=0.03$. They are compatible
with any molecule, provided that the field amplitude is adequately
chosen. For LiCl, for instance, this amounts to a pulse duration of
about 0.3 ps and a field amplitude of $1.5\times 10^5\
\mathrm{V}\,\mathrm{cm}^{-1}$ 
\cite{dion2} in the case of the orientation. Time is indicated in
fractions of the molecular rotational period. The presentation of the
results follows the different steps of the control
strategy. Analytical estimations of the dynamical parameters are
reported in Appendix \ref{appd}.

\subsection{The finite dimensional subspace} \label{section5a}
\begin{figure} 
\includegraphics[width=0.45\textwidth]{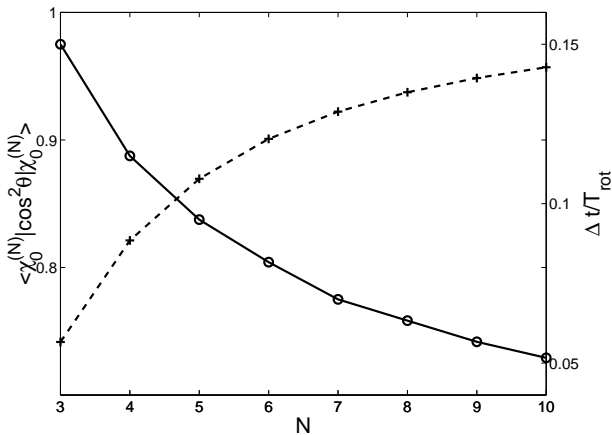}
\caption{\label{fig1} Maximum alignment efficiency (crosses) and
  associated duration (open circles) as a function of $N$, the
  dimension of the rotationally excited subspace
  $\mathcal{H}_{m=0}^{(N)}$ (see text). The solid and dashed lines are
  just to guide the eye.}
\end{figure}

\begin{figure} 
\includegraphics[width=0.45\textwidth]{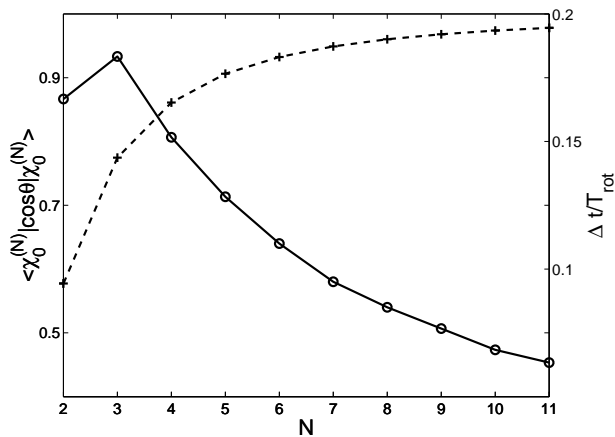}
\caption{\label{fig2} Same as Fig \ref{fig1}, but for orientation.}
\end{figure}

Relevant information for the characterization of alignment and
orientation in the finite subspace $\mathcal{H}_{m=0}^{(N)}$ as a
function of its dimensionality $N$ are displayed in Figs.~\ref{fig1}
and \ref{fig2}.
The maximum efficiency (approximately given by Eq.~(\ref{eq49}) for
the case of orientation) that can ideally be expected for a process
that stays confined within $\mathcal{H}_{m=0}^{(N)}$ is given as
$\langle \chi_0^{(N)}
|\mathcal{O}_{a;o}^{(N)}|\chi_0^{(N)}\rangle$. The maximum duration,
measuring the relative duration of the alignment / orientation
processes over which $\langle \mathcal{O}_{a;o}^{(N)}(s)\rangle$
remains larger than 0.5 during the field-free evolution, is indicated
in terms of a fraction of the rotational period as $\Delta
t/T_{rot}$. The information displayed is crucial for the choice of the
dimensionality $N$ of the reduced subspace: a larger $N$ is more
suitable for a better efficiency of alignment / orientation, but leads
to a shorter duration. The compromise between maximum efficiency and
duration is to be done at that step. In order to keep an alignment /
orientation duration of the order of one tenth of the rotational
period (amounting to durations exceeding 10 ps for heavy diatomic
molecules such as NaI), $N$ has to be limited to 5, as can be observed
from Figs.~\ref{fig1} and \ref{fig2}. This may seem rather limiting in
view of the moderate molecular rotational excitation, but even such a
low dimension ($N=5$) turns out to be sufficient for very efficient
alignment (about 0.85) or orientation (about 0.9). This fixes
completely the target states (approximately given by Eq. (\ref{eq48})
in the case of orientation) for both processes.

\subsection{Dynamics controlled by the two strategies} \label{section5b}

\begin{figure} 
\includegraphics[width=0.45\textwidth]{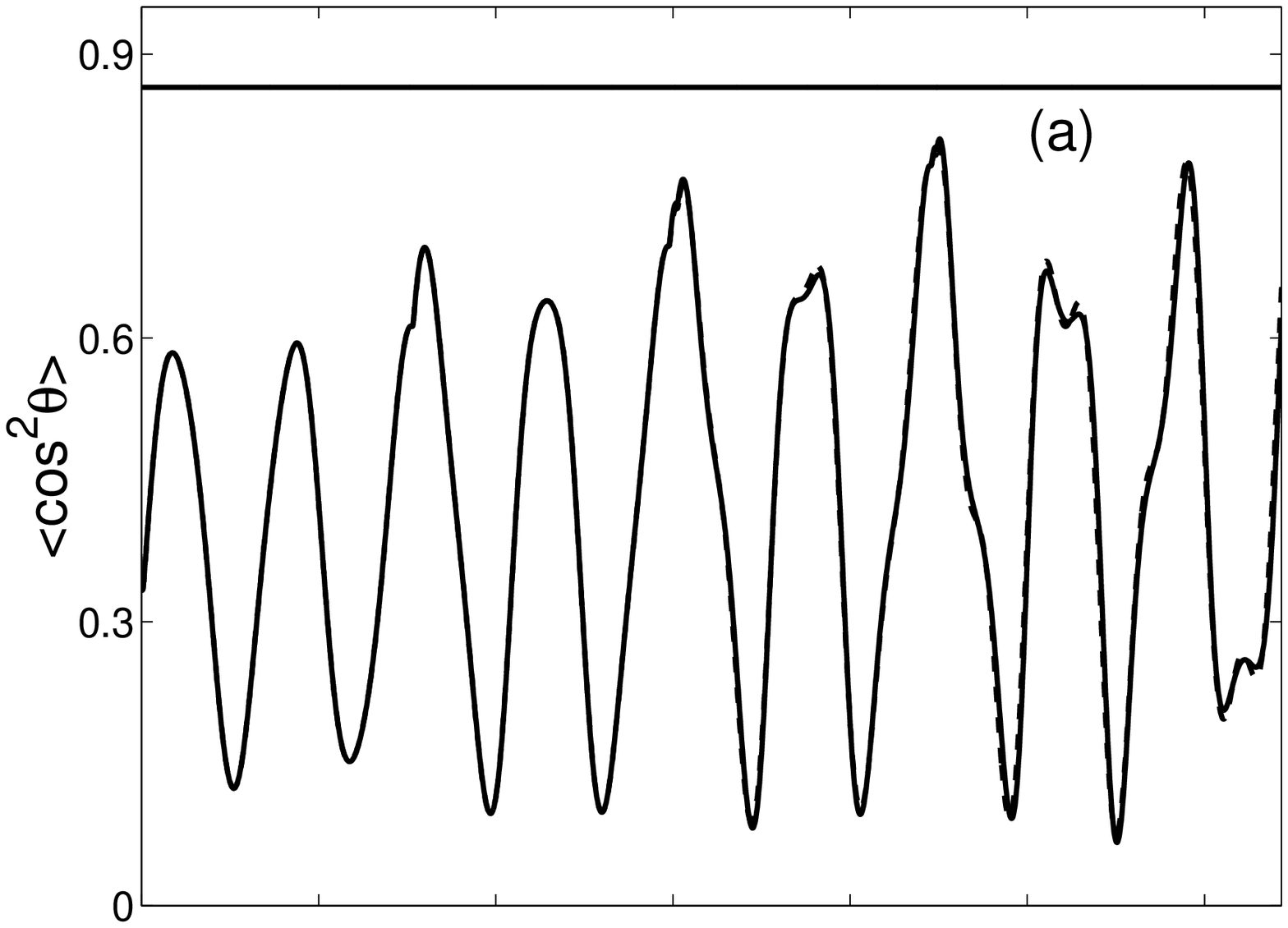}
\includegraphics[width=0.45\textwidth]{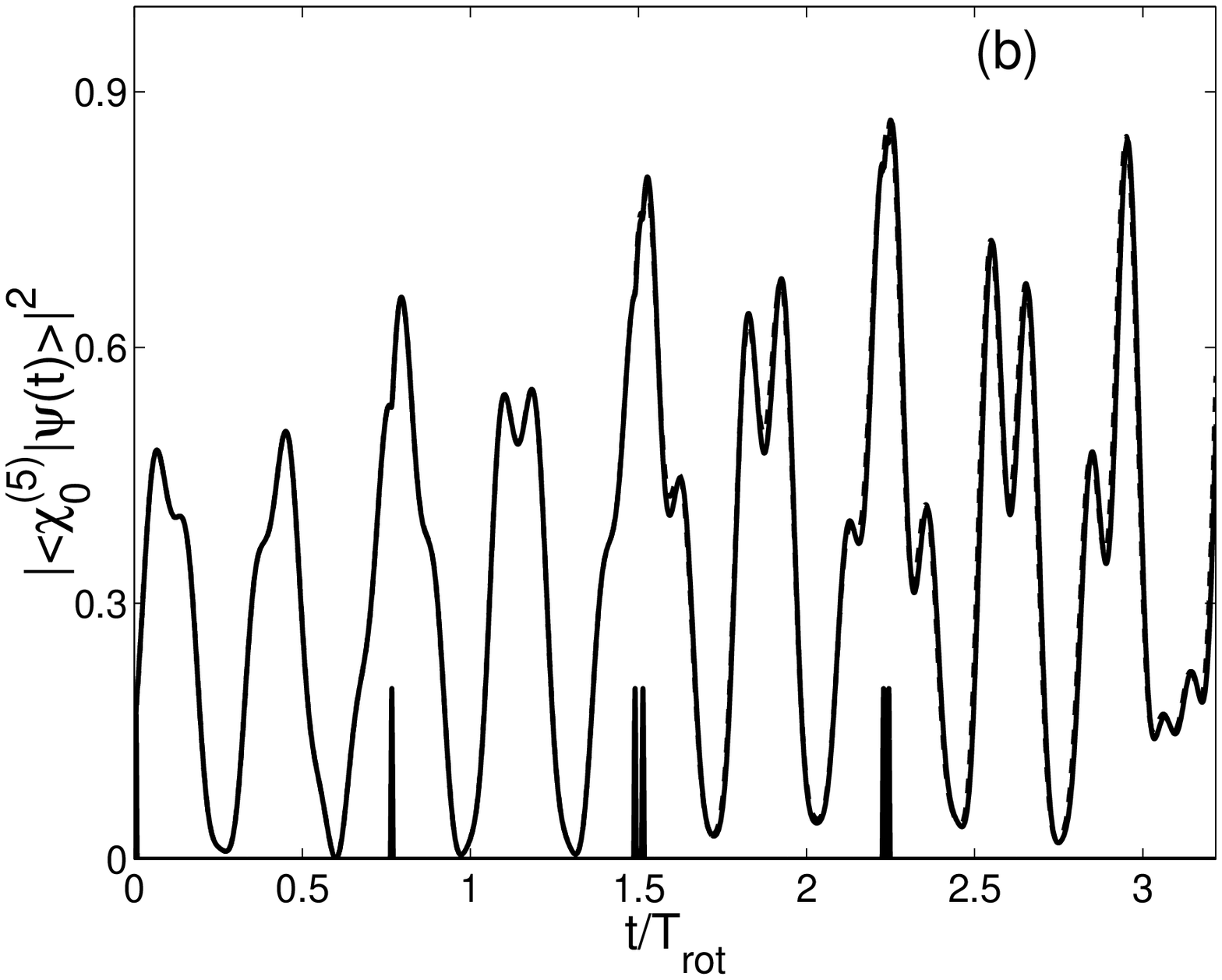}
\caption{\label{fig3} Alignment dynamics during and after the train of
  pulses determined by strategy $S1$ of perturbing at the global
  maxima of $\langle \cos ^2\theta\rangle$ : panel (a) for
  $\langle\psi_\lambda (s)|\cos^2\theta|\psi_\lambda (s)\rangle$ and
  panel (b) for $|\langle\chi_a^{(5)} |\psi_\lambda
  (s)\rangle|^2$. The solid line corresponds to the exactly propagated
  wave function $\psi _\lambda (s)$ and the dashed line to the
  propagation of $\psi _\lambda(s)$ in the subspace
  $\mathcal{H}_{m=0}^{(5)}$. The train of pulses is displayed on panel
  (b) and the optimal alignment in $\mathcal{H}_{m=0}^{(5)}$ is
  indicated by the horizontal line on panel $(a)$.}
\end{figure}

\begin{figure} 
\includegraphics[width=0.45\textwidth]{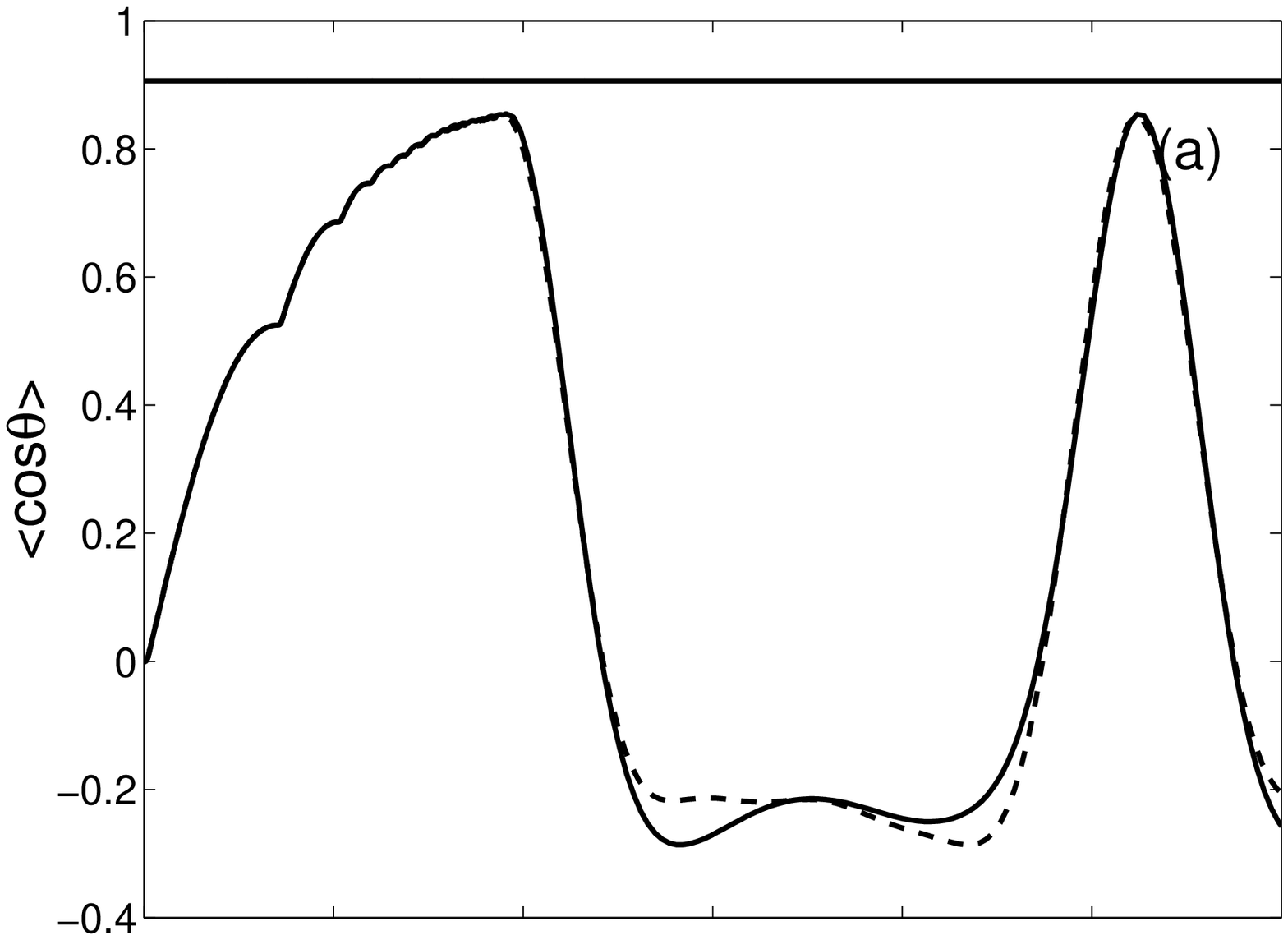}
\includegraphics[width=0.45\textwidth]{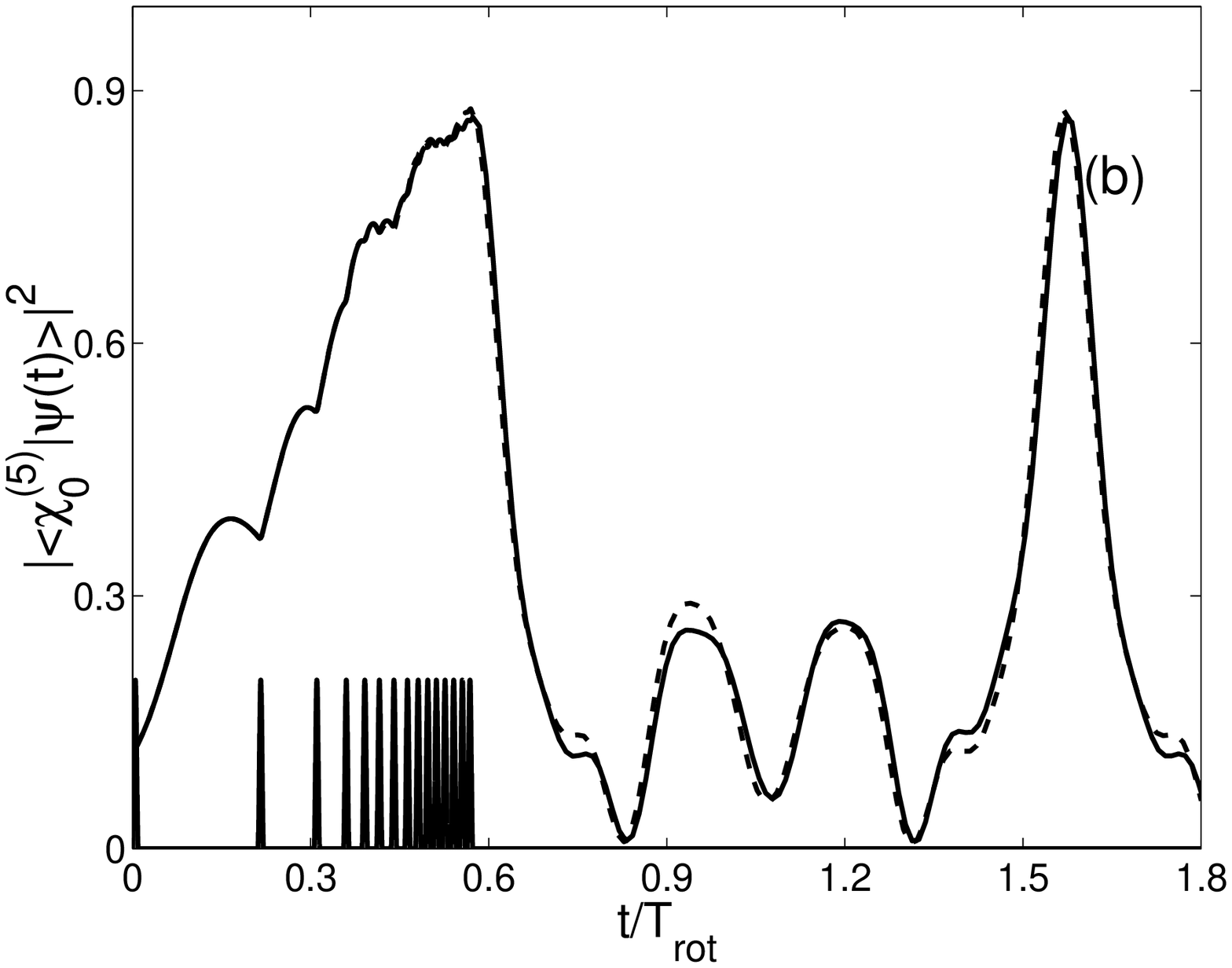}
\caption{\label{fig4} Same as Fig. \ref{fig3}, but for the orientation dynamics.}
\end{figure}

The two strategies, noted $S1$ and $S2$, can be exploited in the same
way, the time delays between the successive pulses being defined
following one of the two processes (maxima of
$\langle\mathcal{O}_{a;o}^{(5)}\rangle$ for $S1$, or maxima of
$|\langle \chi_{a;o}^{(5)} |\psi_\lambda (t)\rangle |^2$ for $S2$).
We look for the dynamical behavior of the two quantities, namely the
expectation value of the observables $\langle
\mathcal{O}_{a;o}^{(5)}\rangle$ and the projections on the target
states $|\langle \chi_{a;o}^{(5)}|\psi_\lambda(t)\rangle|^2$. All
results are displayed with the first pulse taken as the origin of time
and extend at least one rotational period $T_{rot}$ after the last
pulse. This helps to show the complete field-free behavior of the
dynamics, which repeats periodically leading to revival
structures. Figures \ref{fig3} and \ref{fig4} illustrate the results
of applying strategy $S1$ to the alignment and orientation processes,
respectively. Noticeably different behavior is obtained for the two
processes: clearly, a smaller number of pulses is necessary to control
alignment (i.e., 6 pulses for alignment instead of 15 for
orientation).
\begin{figure} 
\includegraphics[width=0.45\textwidth]{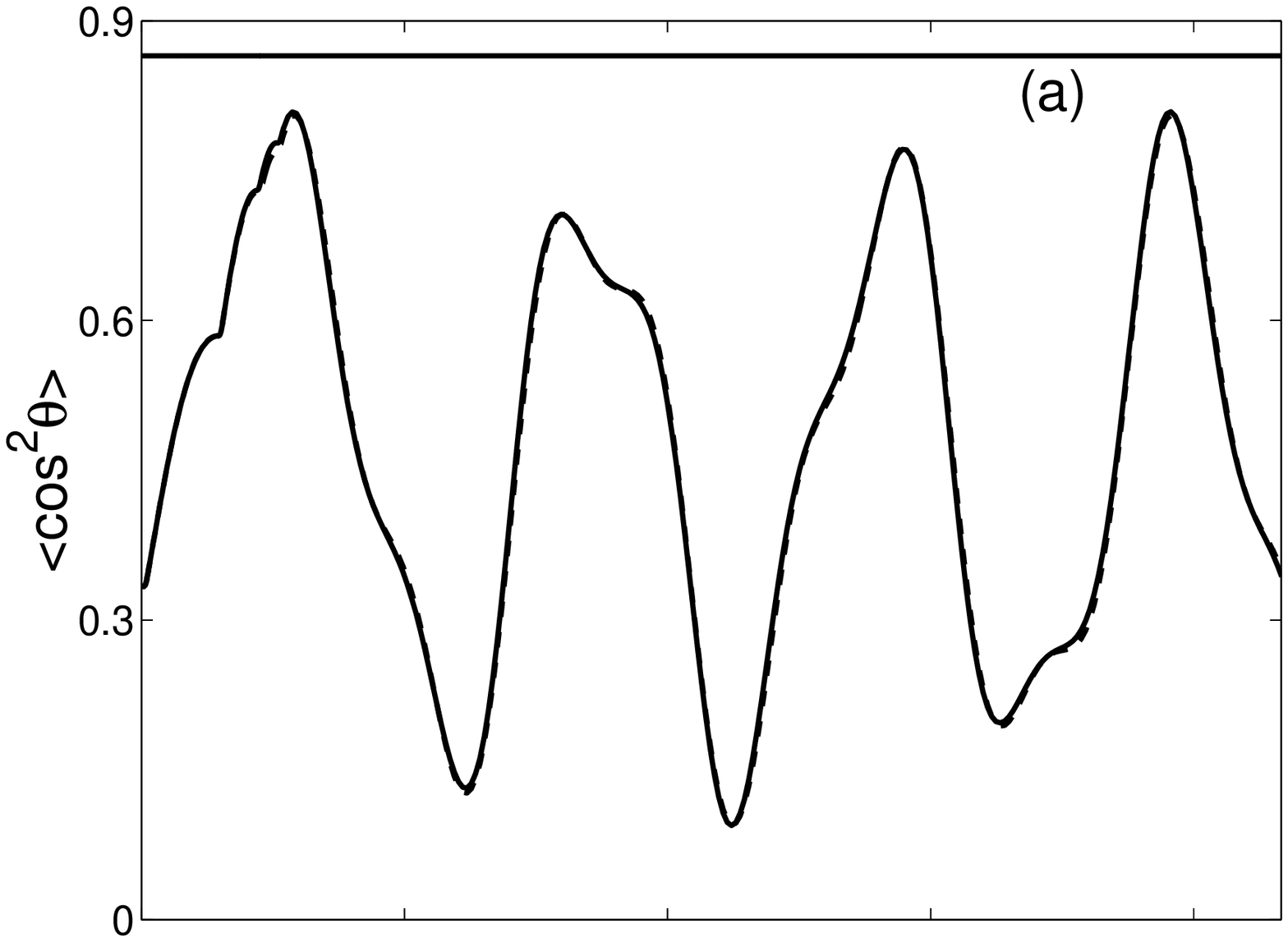}
\includegraphics[width=0.45\textwidth]{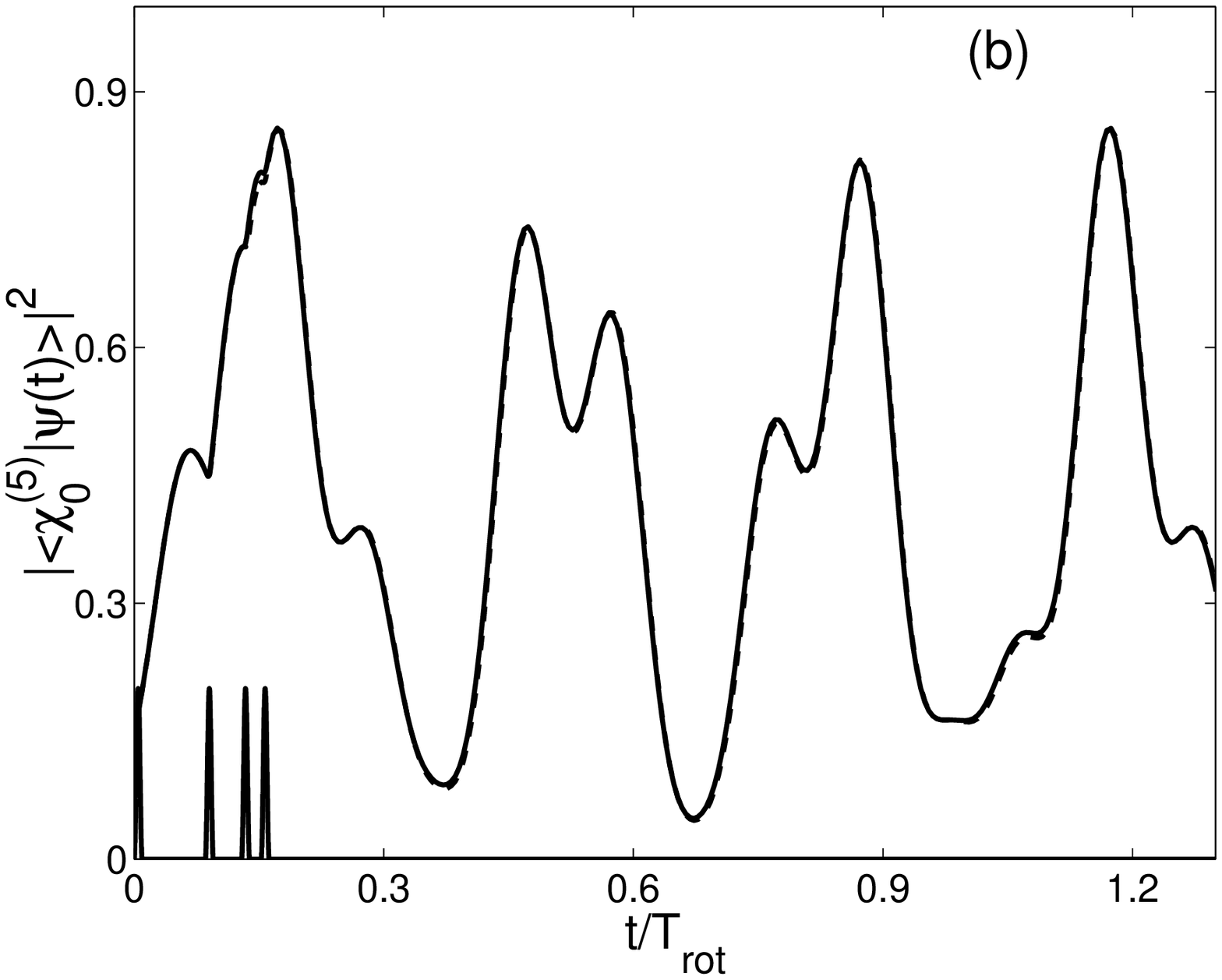}
\caption{\label{fig4a} Same as Fig. \ref{fig3}, but local maxima (the first maximum after the pulse) have been used.}
\end{figure}
The first pulse already provides an alignment efficiency of 0.75,
whereas it only yields 0.5 orientation efficiency. Conversely, the
overall dynamics (conditioning the delays between successive pulses)
is much more oscillatory with complex structures in the case of
alignment as compared to orientation where it progressively increases
up to 0.89, which is almost the optimal limit as found from
Fig.~\ref{fig2}. In all cases, a comparison of the observables and the
projections calculated using the exact wave function $\psi_\lambda(s)$
with the one propagated in the subspace $\mathcal{H}_{m=0}^{(5)}$
supports the claim that the rotational dynamics actually resides
within $\mathcal{H}_0^{(5)}$. In both cases, panels (b) shows the way
the wave function $\psi_\lambda (s)$ gets close to the target state
$\chi _{a;o}^{(5)}$, illustrating the successful outcome of the
control strategy with a coherent choice of parameters $A$, $N$, and
the number of pulses for appropriately describing the
dynamics. Finally, the post-pulse dynamics leads to results
particularly remarkable with respect to previous proposals. An
efficiency of about 0.85 with a duration of (1/10)th of the rotational
period is achieved for alignment, with even better results for
orientation, i.e., 0.89 efficiency and (2/10)th of the rotational
period duration (which corresponds to about 2 ps for a light molecule
like LiCl and 20 ps for a heavy one like NaI).

We next notice a series of local maxima in Fig.~\ref{fig3} (for
instance, just after the first kick), which could be used for the
control strategy. This point is illustrated in Fig.~\ref{fig4a} for
strategy $S1$. It turns out that excellent results are obtained for
the alignment process using in the control scheme the local maxima
(the next one after the application of the pulse). More unexpectedly,
this leads to a better result than the use of the global maxima, as
optimal alignment is almost achieved after only 4 kicks (while 6 are
used in Fig.~\ref{fig3}). In this example, it is clear that the best
choice for the convergence of the process is to kick at a local
maximum. However, we did not manage to establish general rules for
this degree of freedom and it thus seems that numerical tests have to
be undertaken for each practical case.
\begin{figure} 
\includegraphics[width=0.45\textwidth]{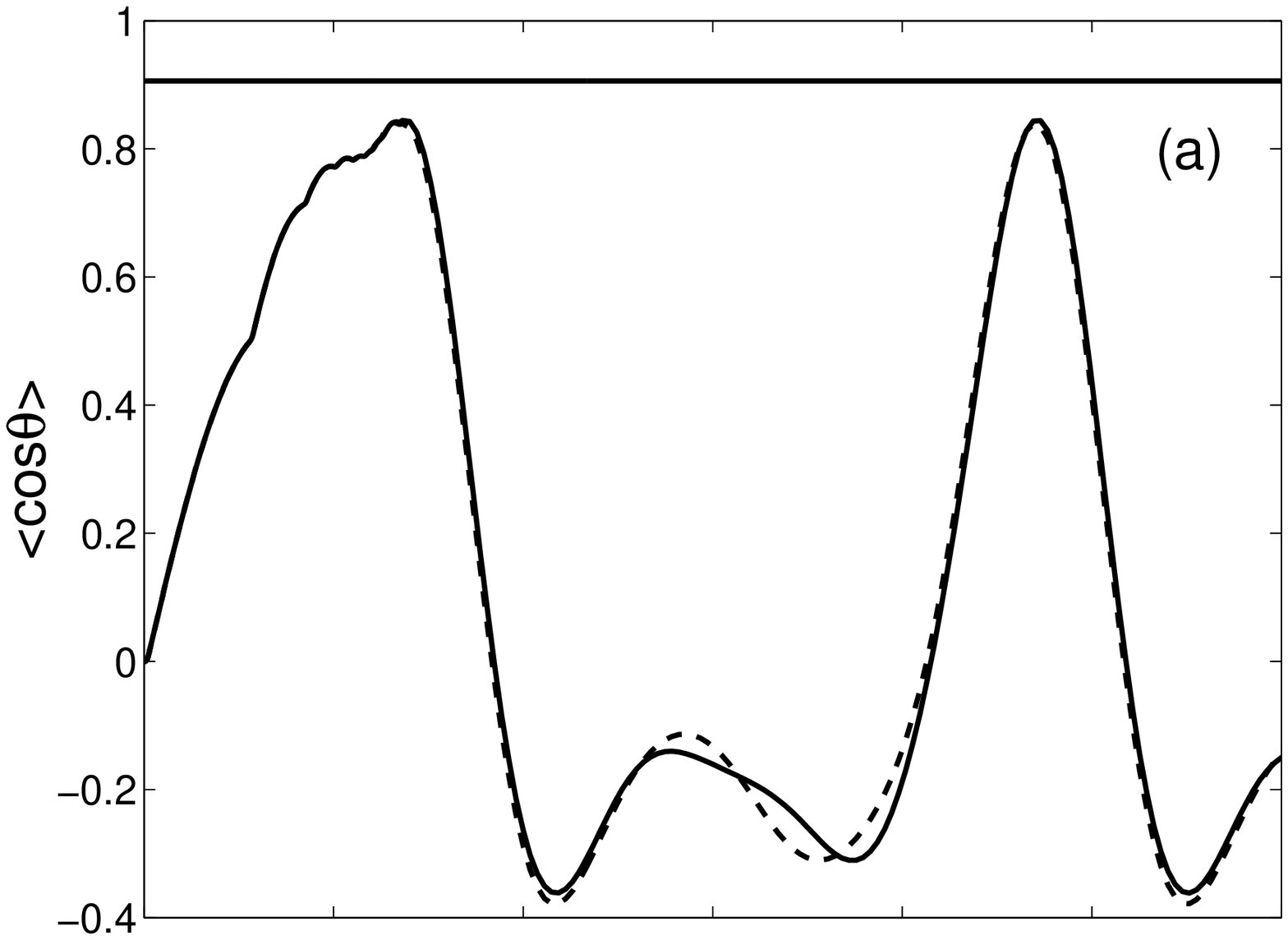}
\includegraphics[width=0.45\textwidth]{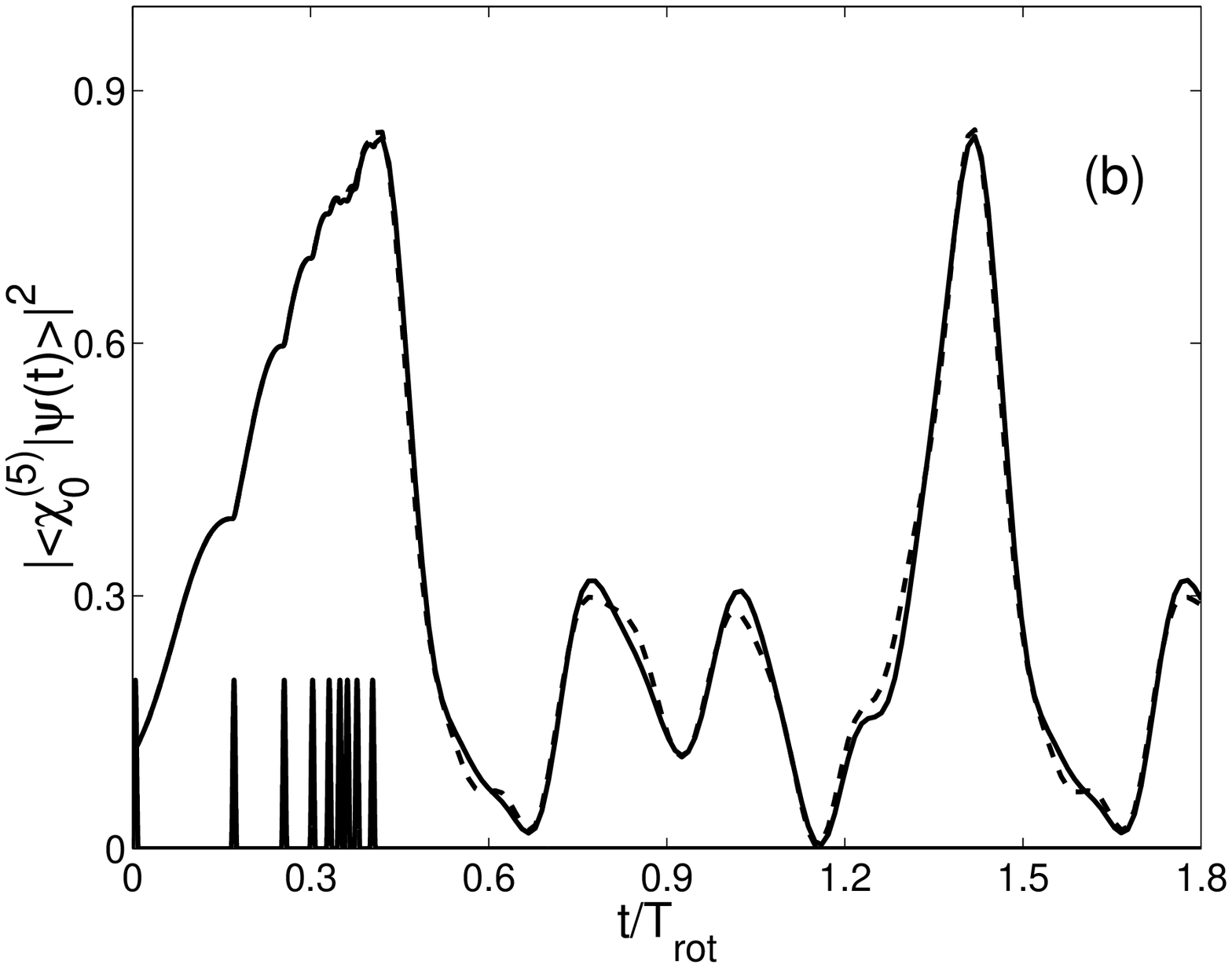}
\caption{\label{fig5} Same as Fig. \ref{fig4}, but referring to
  strategy $S2$, perturbing the system at the maxima of $|\langle
  \chi_o ^{(5)} |\psi_\lambda (s)\rangle |^2$.}
\end{figure}

The outcome of the calculation when applying strategy $S2$ to the
orientation, taken as an example, is displayed in Fig.~\ref{fig5}. The
results obtained are very similar to those using strategy $S1$, with a
smaller number of pulses (viz., 9) but leading to slightly less
efficient orientation. This suggests a great similarity between the
two strategies from a numerical point of view, as can be seen when
comparing Figs. \ref{fig4} and \ref{fig5}.

\subsection{Robustness} \label{section5c}

The robustness of the overall strategy has in principle to be checked
against two variables: the time delays between successive pulses and
the total energy delivered by individual pulses.
\begin{figure} 
\includegraphics[width=0.45\textwidth]{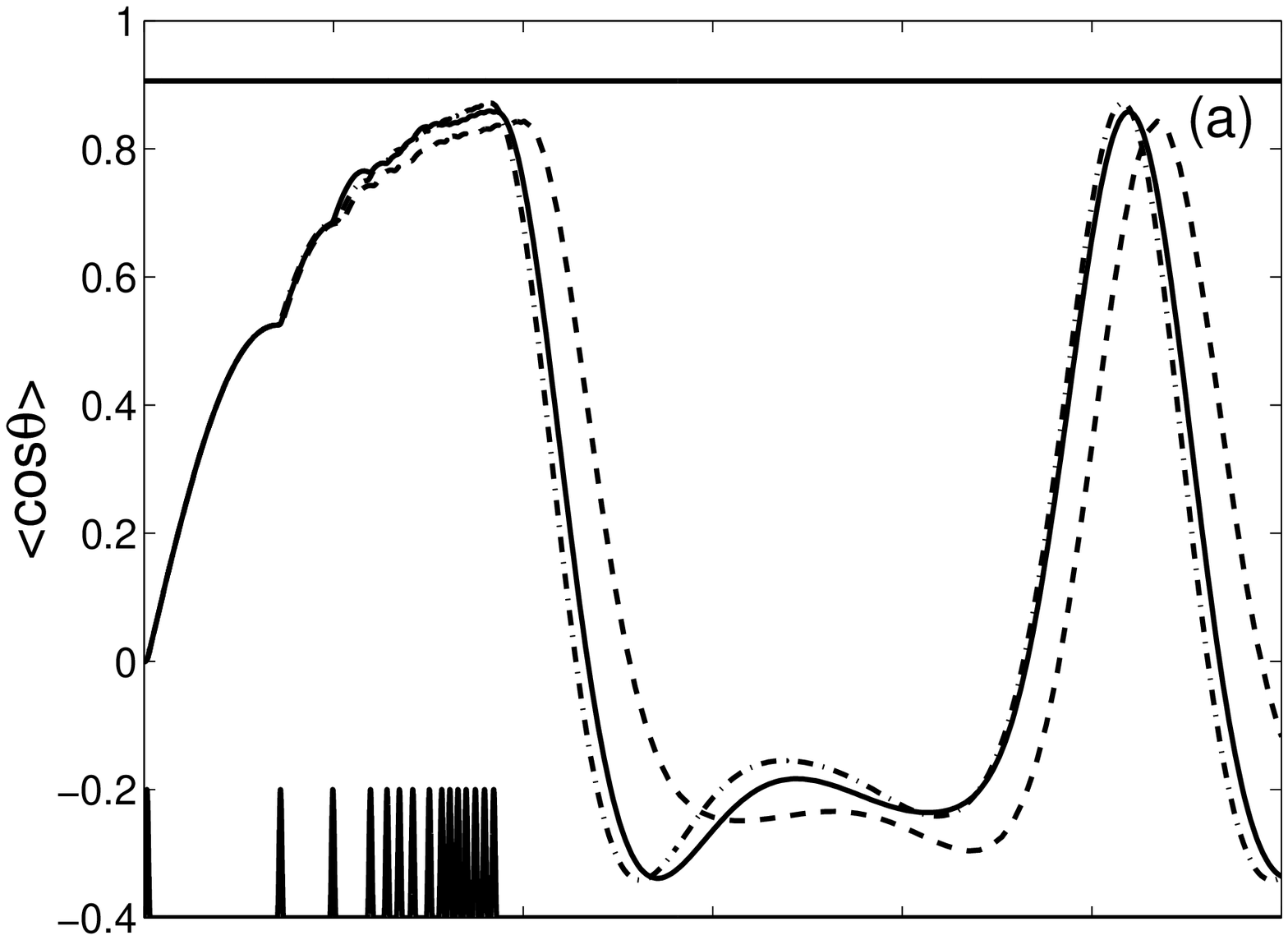}
\includegraphics[width=0.45\textwidth]{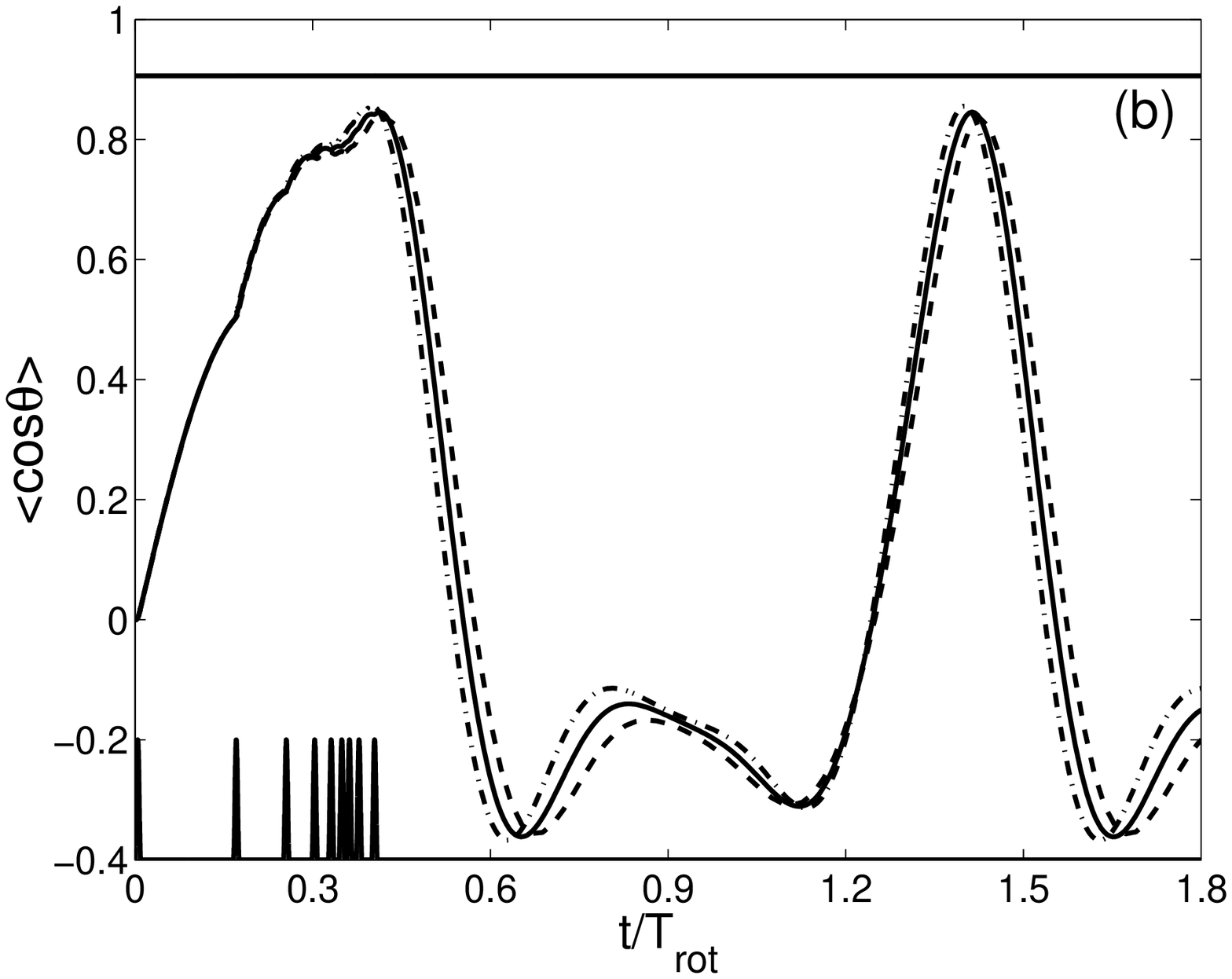}
\caption{\label{fig6} Robustness of the orientation dynamics when
  shifting the time delays between pulses systematically by $\pm 0.1
  \%$ of the rotational period (the dashed line corresponds to
  overestimates by $+0.1\%$ whereas the dotted-dashed one to
  underestimates by $-0.1\%$, the solid line corresponding to the
  calculation of Fig.~\ref{fig4}). The strategy $S1$ is displayed on
  panel (a) and $S2$ on panel (b).}
\end{figure}
Figure \ref{fig6} illustrates the robustness against the most
sensitive parameter, i.e., the time delays $s_i$, where $\langle
\mathcal{O}_o^{(N)}(s_i)\rangle $ is maximum. This is given for the
orientation process only, by considering again the two strategies $S1$
and $S2$. Strategy $S2$ seems here more robust, but this is only due
to the smaller number of pulses which are necessary for reaching the
target state (9 for $S2$, instead of 15 for $S1$).

As for experimental feasibility, the accuracy of the light paths
followed by successive pulses is of the order of $0.2\ \mu
\mathrm{m}$, corresponding to 0.6 fs. This is, for instance for a
light diatomic molecule like LiCl, less than $1/10$th of the
inaccuracies considered when varying the time delays, in
Fig.~\ref{fig6}. In other words, both strategies are very robust with
respect to such inaccuracies affecting the molecule-pulse interaction terms.
\begin{figure} 
\includegraphics[width=0.45\textwidth]{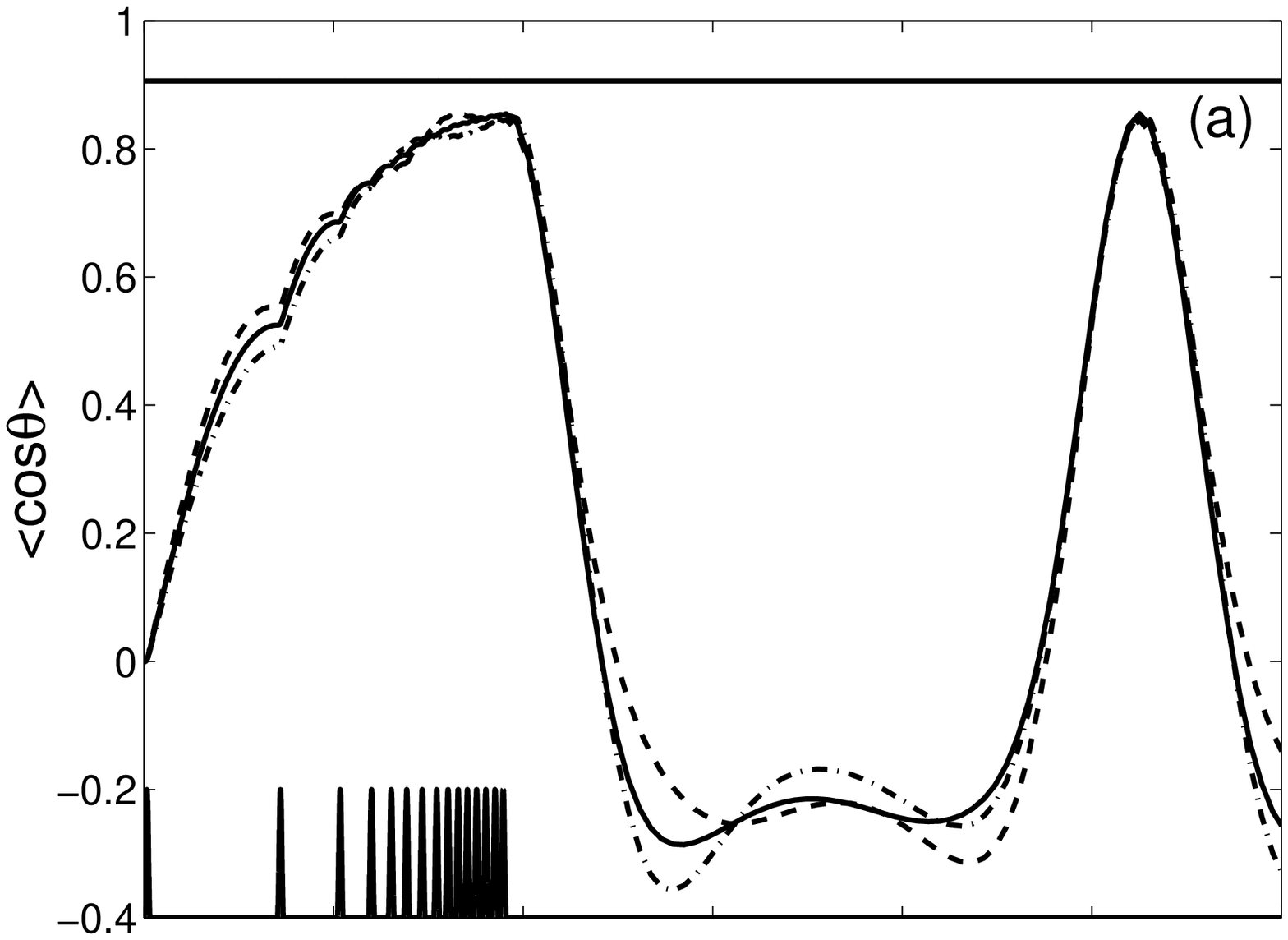}
\includegraphics[width=0.45\textwidth]{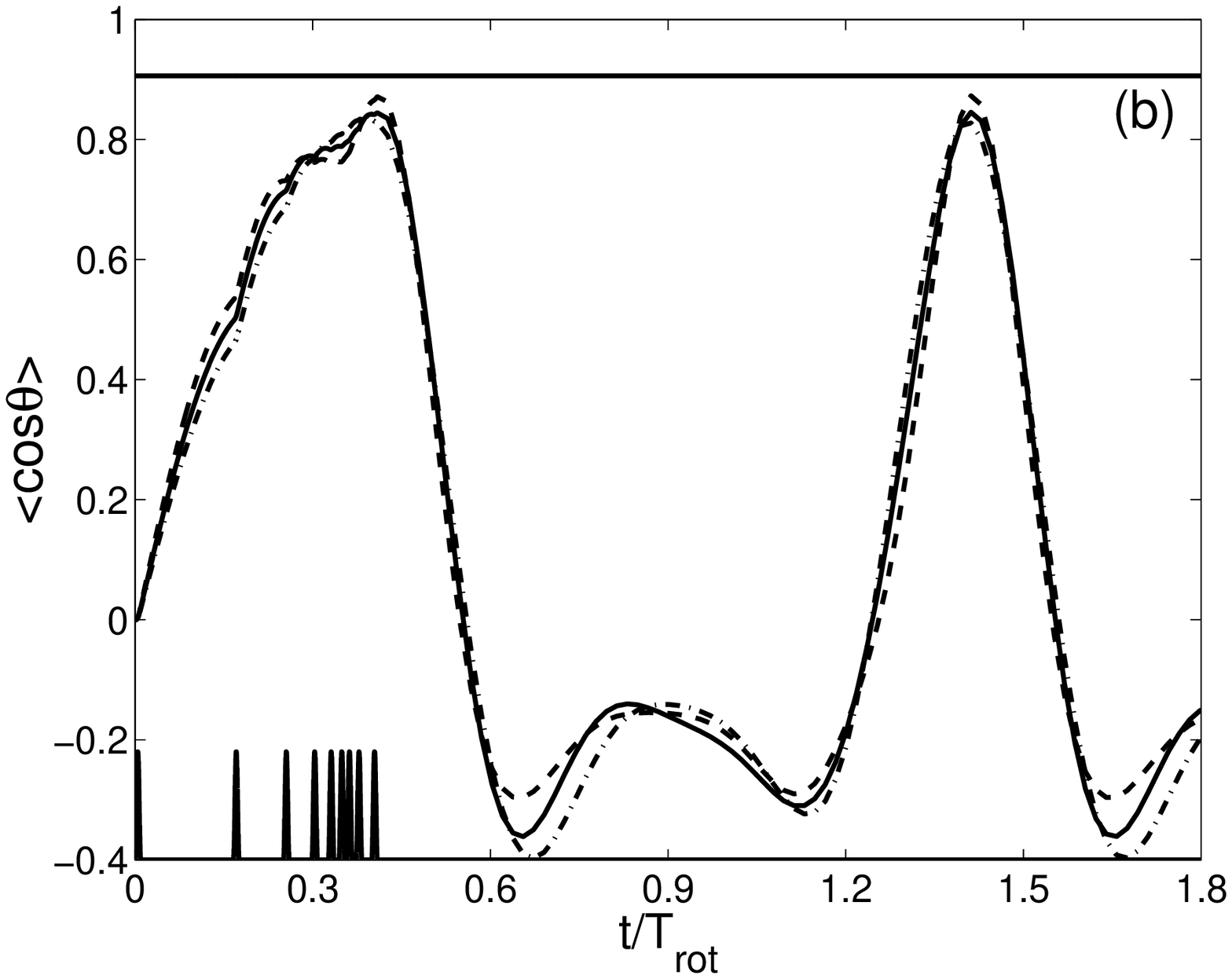}
\caption{\label{fig7} Robustness of the orientation dynamics varying
  the pulse energy by $10\%$. The full line corresponds to $A=1$,
  i.e., the calculation of Fig.~\ref{fig4}, the dotted and
  dotted-dashed lines are for $A=1.1$ and $A=0.9$, respectively. The
  strategy $S1$ is displayed on panel (a) and $S2$ on panel (b).}
\end{figure}
On the contrary, the measure of the total energy per pulse is affected
by a more severe inaccuracy. This is checked in Fig.~\ref{fig7}, by
varying $A$ by 10$\%$, for both strategies $S1$ and $S2$. Remarkable
robustness is achieved, advocating for an experimental feasibility of
the control scheme: a few short pulses (say 5 to 6), with a moderate
total energy determined within 10$\%$ of accuracy, applied with time
delays of the order of 1 ps with an accuracy of 1 fs, leads to an
excellent control of the orientation for any molecular system.

\section{Conclusions} \label{section6}

In this work, we have developed a laser control strategy for a
molecular system initially in a pure quantum state, aiming at the
maximalization (or minimalization) of an observable $\mathcal{O}$ given by
an operator, not commuting with the molecular field-free
Hamiltonian. The strategy consists of four steps: reduction of the
dimensionality of the original physical Hilbert space; definition of a
target state as an eigenvector of the reduced observable in the finite
subspace; analysis of the controllability of the system; application
of a train of sudden perturbations $U$, commuting with $\mathcal{O}$,
at times where the expectation value of $\mathcal{O}$ reaches a
maximum during its field-free evolution. A process similar in its
spirit consists in applying the pulses at times when the projection of
the time-evolved wave function on the target states reaches maximum
values. This latter strategy yields comparable numerical results even
if, from a theoretical point of view, its set of fixed points (which
corresponds only to the target state) can be smaller than (or equal to) the one of the
first process. Moreover, it is worth noting that such general
strategies are not only independent of the particular molecule, but
are also transferable to different control issues and different
observables. Finally, we have checked the robustness with respect to
inaccuracies in the time delays between pulses or the total
pulse energies, and found it to be remarkably good, as long as
moderate intensity and short radiative interactions are considered.

Control strategies based on pure states are however subject to
important modifications when dealing with temperature
effects. Orientation, for instance, drastically decreases for
increasing temperatures
\cite{machholm,machholm1,ortigoso,benhajyedder}. This is basically due
to the fact that, at non-zero temperature, the initial state is in
general not a pure state, but a mixed one, i.e., a superposition of a
statistical ensemble of rotational states with $m\neq 0$, which tend
to misalign the molecule. The dynamics of such mixed states is
described by a density operator $\rho (t)$ (instead of a wave
function) evolving according to the von Neuman equation (instead of
the Schr\"{o}dinger equation). Following the dimensionality reduction of
the Hilbert space, the control strategy looks now for the optimal
density operator $\rho ^{(N)}$ which maximizes the observable
$\mathcal{O}^{(N)}$ in the same subspace. This amounts to
maximizing $Tr[\rho ^{(N)}\mathcal{O}^{(N)}]$, with the
commutation relation $[\rho ^{(N)},\mathcal{O}^{(N)}]=0$. It has been
shown \cite{girardeau,girardeau1} that such a goal can be achieved by
an unitary operator $U$ that appropriately orders the common
eigenvectors of $\rho ^{(N)}$ and $\mathcal{O}^{(N)}$, such that the
one corresponding to the highest eigenvalue of $\rho ^{(N)}$ be
associated with the one corresponding to the highest eigenvalue of
$\mathcal{O}^{(N)}$, and so forth. An even more sophisticated
generalization of the control strategy is to be looked for when the
particular system under consideration evolves while interacting with a
physical environment describing collisional processes in a gas, or
friction in a liquid \cite{shlomo,tannor}.  The strategies that have
to be worked out would involve non-unitary perturbations $U$ and a
special attention has to be paid to decoherence
\cite{beige,lidar}. Work in these directions is in progress.
\begin{acknowledgments}
  Support from the Conseil R\'{e}gional de Bourgogne is gratefully
  acknowledged.
\end{acknowledgments}

\appendix
\section{Analysis of the set of fixed points: the general
  case}\label{appa} 
The goal of this appendix is to determine the set of fixed points of
the control strategy $S1$. Indeed, as mentioned in
Sec. \ref{section2}, a crucial question for the convergence of the
control scheme is the set $\mathcal{F}$ of its fixed points. In
general, we are not able to exactly determine $\mathcal{F}$ but only a
larger set $\mathcal{S}$. The latter contains the wave functions
$|\psi_s\rangle$ which fulfill the requirements
\begin{equation} \label{eq4b}
\langle \psi_s|[H_0,\mathcal{O}^{(N)}]|\psi_s\rangle =0 
\end{equation}
and
\begin{equation} \label{eq4c}
\langle
\psi_s|U_{\tilde{A}}^{-1}[H_0,\mathcal{O}^{(N)}]U_{\tilde{A}}|\psi_s\rangle
=0 , 
\end{equation}
for all values of the real parameter $\tilde{A}$. As mentioned above,
we can show that $\mathcal{F}$ is a subset of $\mathcal{S}$. For the
purpose of proving this result, we observe that just before the pulse
at $t_i-0$ one has
\begin{equation} \label{eq12}
\left. \frac{d}{dt} \langle \mathcal{O}^{(N)} \rangle \right|_{t_i-0}
= \left. i\langle [H_0,\mathcal{O}^{(N)}]\rangle \right|_{t_i-0}=0  ,
\end{equation}
whereas just after the pulse at $t_i+0$ 
\begin{eqnarray} \label{eq13}
\left. \frac{d}{dt} \langle \mathcal{O}^{(N)} \rangle \right|_{t_i+0}
&=& \left. i\langle
  U_{\tilde{A}}^{-1}[H_0,\mathcal{O}^{(N)}]U_{\tilde{A}}\rangle
\right|_{t_i+0}  \\  
& = & \left. i\langle
  [U_{\tilde{A}}^{-1}H_0U_{\tilde{A}},\mathcal{O}^{(N)}]\rangle  \right|_{t_i+0} .
\end{eqnarray}
We can conclude by arguing that if there exists a value of the
parameter $\tilde{A}$ such that the slope undergoes a change from zero
[Eq.~(\ref{eq12})] to a non-zero finite value [Eq.~(\ref{eq13})], then
$\langle\mathcal{O}^{(N)}(t)\rangle$ will reach within the period $T$
a maximum strictly larger than the one obtained prior the application
of the pulse.

We are now in a position to state the following theorem:
$\mathcal{S}=\{ |\chi^{(n)}\rangle\}$ (the $|\chi^{(n)}\rangle$'s
being the eigenstates of $\mathcal{O}^{(N)}$) if and only if the
dimension of the vector space $\mathcal{V}$ generated by
$[H_0,\mathcal{O}^{(N)}]$ and the elements of the form
$U_{\tilde{A}}^{-1}[H_0,\mathcal{O}^{(N)}]U_{\tilde{A}}$, where
$\tilde{A}\in\mathbb{R}$, is equal to $N(N-1)$. The proof goes as
follows.

We first recall that the complete controllability of the system is
assumed. This condition is equivalent to requiring that the Lie
algebra $\mathcal{L}$ generated by $iH_0$ and $iH_I$ is $u(N)$ or that
the dimension of this vector space is $N^2$. Moreover,
$U_{\tilde{A}}^{-1}[H_0,\mathcal{O}^{(N)}]U_{\tilde{A}}$ being a
skew-hermitian operator, it can be shown that $\mathcal{V}$ is a
subspace of $\mathcal{L}$.
The starting point for the following computations will be the standard
basis of $u(N)$ \cite{fu},
\begin{eqnarray} \label{eq4d}
\left\{
\begin{array}{lll}
x_{mn} &=& e_{mn}-e_{nm}, \\
y_{mn} &=& i(e_{mn}+e_{nm}), \\
d_{nn} &=& ie_{nn},
\end{array} \right. 
\end{eqnarray}
where $e_{mn}=|\chi^{(m)}\rangle\langle \chi^{(n)}|$, $1\leq n\leq
N-1$ and $n<m \leq N$. We denote $D$ the dimension of
$\mathcal{V}$. This vector space is generated by the set $v_k$
($k=1,\ldots,D$), which can be written as a linear combination of
elements of the basis given by Eq.~(\ref{eq4d}). Expanding
$[H_0,\mathcal{O}^{(N)}]$ and
$U_{\tilde{A}}^{-1}[H_0,\mathcal{O}^{(N)}]U_{\tilde{A}}$ with respect
to the previous set $v_k$ and taking the mean value of these operators
over a wave function $|\psi_s\rangle \in\mathcal{S}$, we obtain a system of
linear equations. Inverting the system, one gets $D$ equations of the
form
\begin{equation} \label{eq4e}
\langle \psi_s|v_k|\psi_s\rangle =0 .
\end{equation}
As $|\chi^{(n)}\rangle\in\mathcal{S}$ and $\langle
\chi^{(n)}|d_{nn}|\chi^{(n)}\rangle=1$ for any $n=1,\ldots,N$, it is
clear that $\mathcal{V}$ does not contain any linear combination of
the $d_{nn}$'s, which shows that the dimension of $\mathcal{V}$ is at
most $N(N-1)$. Hence, if we assume that, for instance, $x_{mn}\in
\mathcal{V}$ and $y_{mn}\in\mathcal{V}$, we obtain by introducing the
wave function $|\psi\rangle=\sum_{n=1}^N c_n|\chi^{(n)}\rangle$ into
Eq.~(\ref{eq4e})
\begin{eqnarray} \label{eq4f}
\left\{ \begin{array}{ll}
c_m^*c_n-c_n^*c_m = 0, \\
c_m^*c_n+c_n^*c_m = 0 .
\end{array}\right. 
\end{eqnarray}
From Eqs.~(\ref{eq4f}), one deduces that $c_m=0$ or $c_n=0$. By
repeating this procedure for each couple $(m,n)$, it can be shown that
the states $|\chi^{(n)}\rangle$, for which for some $n_0$
\begin{eqnarray} \label{eq4g}
\left\{ \begin{array}{ll}
c_{n_0} =1, \\
c_{m;m\neq n_0} = 0,
\end{array}\right.
\end{eqnarray}
are the unique elements of $\mathcal{S}$ if and only if the dimension
of $\mathcal{V}$ is $N(N-1)$. Numerical calculations of this dimension
and complementary results are presented in Appendix
\ref{appb}. Finally, a conclusion regarding this theorem is that if
the dimension of $\mathcal{V}$ is $N(N-1)$, then the set $\mathcal{F}$
of fixed points of the control strategy is a subset of
$\{|\chi^{(n)}\rangle\}$, the set of eigenvectors of the impulsive
propagator $U_{\tilde{A}}$, which are also the ones of
$\mathcal{O}^{(N)}$ as the two operators commute.

\section{Analysis of the set of fixed points in the sudden
  approximation} \label{appb} 

This appendix particularizes the general arguments of the Appendix
\ref{appa} to the sudden approximation.  We first consider the
strategy $S1$. In this case, we have to determine the dimension of the
vector space $\mathcal{V}$ defined in Appendix \ref{appa}. In the
sudden approximation, we recall that the evolution operator
$U_{\tilde{A}}$ can be written as [Eq.~(\ref{eq11})] 
\begin{equation} \label{eqap0}
U_{\tilde{A}}=e^{i\tilde{A}H_I} ,
\end{equation}   
and $\mathcal{V}$ is then generated by the elements of the form
$e^{-i\tilde{A}H_I}[H_0,\mathcal{O}^{(N)}]e^{i\tilde{A}H_I}$, where
$\tilde{A}\in \mathbb{R}$. For the sake of clarity, we now consider
that $H_I=\mathcal{O}^{(N)}$ (this assumption is fulfilled for the
alignment / orientation processes). Using the Campbell-Hausdorf
formula \cite{merzbacher}, simple algebra shows that $\mathcal{V}$ is
also spanned by the operators $ad^n(H_0,\mathcal{O}^{(N)})$ ($n\geq
1$). The notation $ad^n(B,C)$ for the operators $B$ and $C$ is defined
by the recurrence formula 
\begin{eqnarray}
ad^n(B,C)=
\left\{\begin{array}{ll}
B & n=0 \\
\lbrack ad^{n-1}(B,C),C\rbrack & n\geq 1
\end{array} \right. .
\end{eqnarray}
From a practical point of view, owing to the complexity of analytical
calculations, a numerical algorithm, which closely follows the one for
complete controllability \cite{schirmer2}, can be used to determine
the dimension of $\mathcal{V}$. The first step consists in rewriting
the matrix of each operator $ad^n(H_0,\mathcal{O}^{(N)})$ ($n\geq 1$)
as an $N^2$ column vector obtained by concatenating its columns. We
next construct a new matrix $R$ with these vectors, each vector
corresponding to a column of $R$. Finally, the dimension of
$\mathcal{V}$ is given by the rank of $R$.

We apply this method to the orientation process in a finite
dimensional space. The results of these computations are presented in
Tab.~\ref{tab1}, where the notation $\dim(\mathcal{V})$ corresponds to
the dimension of the vector space $\mathcal{V}$.
\begin{table}[ht]
  \caption{\label{tab1} Dimensions of the vector spaces $\mathcal{L}$
    and $\mathcal{V}$ as a function of $N$ for the orientation
    dynamics. The third column indicates the maximum of the dimension
    of $\mathcal{V}$.} 
  \begin{center} 
\begin{tabular}{c|c|c|c} 
\hline
N & $\dim(\mathcal{L}$) & Max. of $\dim(\mathcal{V}$) & $\dim(\mathcal{V}$) \\
\hline
\hline
3 & 9 & 6 & 4 \\
4 & 16 & 12 & 8 \\
5 & 25 & 20 & 12 \\
\hline 
\end{tabular}
\end{center}
\end{table}
For the orientation, as $\dim(\mathcal{V})<N(N-1)$, one deduces that
$\mathcal{S}\neq\{|\chi^{(n)}\rangle\}$. This point can also be
understood through the following calculation.

We introduce the operator $C=\lbrack H_0,\mathcal{O}^{(N)}\rbrack$ and
we note $C_{mn}$ its matrix elements in the basis
$\{|\chi^{(n)}\rangle\}$,
\begin{equation} \label{app027}
C_{nm}=\langle \chi^{(n)}|C|\chi^{(m)}\rangle .
\end{equation}
For operators $H_0=J^2$ and $\mathcal{O}^{(N)}=\cos^{(N)}\theta$, it
can furthermore be checked that $C_{mn}=0$ if $m=n$ and $C_{mn}\neq 0$
otherwise.  Assuming that the slope is zero before and after the
pulse, we can easily show that
\begin{eqnarray} \label{app028}
\left\{ \begin{array}{ll}
\sum_{m,n=1}^N c_m^*c_nC_{mn}=0 , \\
\sum_{m,n=1}^N e^{i\tilde{A}(\chi_m-\chi_n)}c_m^*c_nC_{mn}=0 ,
\end{array} \right. 
\end{eqnarray}
where the $c_n$'s are the weighting coefficients of the wave function
just before the pulse in the basis of the eigenstates of
$\mathcal{O}^{(N)}$, and the $\chi_n$'s the eigenvalues of
$\mathcal{O}^{(N)}$.  From Eqs. (\ref{app028}), one then deduces that
\begin{equation} \label{app029}
\sum_{m\neq n}e^{i\tilde{A}(\chi_m-\chi_n)}c_m^*c_nC_{mn}=\sum_{m\neq
  n}c_m^*c_nC_{mn} .
\end{equation}
To conclude, we will use the following notion. A group of energy
levels of $\mathcal{O}^{(N)}$ is said to be equally spaced if there exists a
set of integers $\{p,q,r,s\}\in\{1,2,\ldots,N\}^4$ such that 
\begin{equation} \label{app030}
\chi_p-\chi_q=\chi_r-\chi_s  .
\end{equation}
We next suppose that the spectrum of $\mathcal{O}^{(N)}$ is not
equally spaced (according to the previous definition) and that the
coefficients $C_{mn;m\neq 0}\neq 0$. Using the fact that the
left-hand-side of Eq.~(\ref{app029}) is a regular function of the
parameter $\tilde{A}$ that is constant if and only if all its
coefficients $c_m^*c_nC_{mn}$ are zero, we see that the unique
solutions of Eq.~(\ref{app029}) are the eigenvectors
$|\chi^{(n)}\rangle$ of $\mathcal{O}^{(N)}$ for which
\begin{eqnarray} \label{app031}
\left\{\begin{array}{ll}
c_n=1, \\
c_{m;m\neq n}=0 .
\end{array}
\right. 
\end{eqnarray}
It is interesting to note that the two previous hypothesis also give
practical conditions on $\mathcal{O}^{(N)}$ and $H_0$ in order to
determine the set $\mathcal{S}$. Indeed, for the orientation process,
we notice that the spectrum of $\cos^{(N)}\theta$ being equally spaced
(the spectrum is symmetric), the dimension of $\mathcal{V}$ is not
equal to the maximum dimension $N(N-1)$.
   
Let us now determine the set $\mathcal{F}$ of fixed points for the
strategy $S2$ in the sudden approximation. In this case, we first
clarify the definition of $\mathcal{S}$. It can be shown that
$\mathcal{S}$ contains the wave functions $|\psi\rangle$ which fulfill
\begin{equation} \label{app032}
\Im \left[\langle \chi ^{(N)}|e^{i\tilde{A}\mathcal{O}^{(N)}}|\psi\rangle\langle\psi|e^{-i\tilde{A}\mathcal{O}^{(N)}}H_0|\chi^{(N)}\rangle \right]=0 
\end{equation} 
for any value of $\tilde{A}$, with $\Im$ corresponding to the
imaginary part. Eq.~(\ref{app032}) can be rewritten as
\begin{equation} \label{app032a}
\sum_{m=1}^N \Im \left[c_Nc_m^*h_{mN}e^{-iA(\chi_m-\chi_N)} \right]=0 ,
\end{equation}
the $h_{mn}$'s being the matrix coefficients of $H_0$ in the basis
$\{|\chi^{(n)}\rangle\}$. If the spectrum of $\mathcal{O}^{(N)}$ is
completely non degenerate, one deduces from Eq.~(\ref{app032a}) that
\begin{equation} \label{app033}
\Im \left[c_Nc_m^*h_{mN}e^{-iA(\chi_m-\chi_N)} \right]=0
\end{equation}
for any value of $m$ ($m\neq N$). Simple algebra shows that 
\begin{equation} \label{app034}
c_Nc_m^*h_{mN}=0 .
\end{equation}
If we now assume that $h_{mN}\neq 0$, one obtains that $c_N=0$ or
$c_m=0$. The case $c_N=0$ corresponds to the minimum of the projection
$|\langle \psi|\chi^{(N)}\rangle|^2$, which is not relevant
here. Finally, in the other case, we can conclude that the unique
element of $\mathcal{S}$ is $|\chi^{(N)}\rangle$. We also remark that
these two conditions are fulfilled for the orientation process. In
this example, we have therefore shown that the limit of the sequence
is $|\chi^{(N)}\rangle$.

\section{The molecular alignment / orientation control strategy : the
  infinite and the finite dimensional cases} \label{appc}

In this appendix, we shall focus on the convergence of the sequence in
an infinite and a finite dimensional Hilbert space, as only the finite
case has been treated in Appendices \ref{appa} and \ref{appb}.  We
first consider the infinite dimensional case. We will explicitly
derive Eqs.~(\ref{eq550}) and (\ref{eq551}). For doing so, we recall
the following fundamental commutation relations \cite{joyeux}, which
will be used below:
\begin{eqnarray} \label{eqap2}
\begin{array}{ccc}
\lbrack J^2,\cos \theta\rbrack&=& 2(\sigma_\theta+\cos\theta), \\ 
\lbrack \sigma_\theta,\cos\theta\rbrack &=& \cos^2\theta-\textbf{1}, \\
\lbrack J^2,\cos^2 \theta\rbrack &=& 4\cos\theta \sigma+6\cos ^2 \theta -2, \\
\lbrack \sigma,\cos ^2 \theta\rbrack &=& 2 \cos ^3 \theta-2\cos\theta, \\
\lbrack \cos\theta \sigma_\theta,\cos ^2\theta \rbrack &=& 2 \cos
^4\theta-2\cos ^2 \theta, 
\end{array} 
\end{eqnarray}
where $\sigma_\theta =\sin\theta \frac{\partial}{\partial\theta}$. We
notice that such identities are not fulfilled in a finite dimensional
subspace. We consider the strategy $S1$ for the alignment and the
orientation processes. We assume that the average $\langle
\cos^m\theta\rangle$ reaches a maximum at a time $s_i$,
\begin{equation} \label{eqap3}
\left. \frac{d}{ds}\langle \cos ^m\theta \rangle \right|_{s_i-0} =i\langle [\varepsilon J^2,\cos^m \theta] \rangle =0 ,
\end{equation}
where $m=1$ for the orientation and $m=2$ for the alignment. A sudden
pulse being applied at the time $s_i$, we have to calculate the slope
\begin{equation} \label{eqap4}
\left. \frac{d}{ds}\langle \cos ^m\theta \rangle \right|_{s_i+0} =i\langle e^{-iA\cos ^m\theta}\varepsilon [J^2,\cos^m \theta] e^{iA\cos ^m\theta} \rangle .
\end{equation}
Using Eqs.~(\ref{eqap2}) and (\ref{eqap4}), one readily obtains for
the orientation
\begin{equation} \label{eqap5}
\left. \frac{d}{ds}\langle \cos \theta \rangle \right|_{s_i+0}
=i\langle \varepsilon [J^2,\cos \theta]\rangle +2\varepsilon A
(\textbf{1}-\langle\cos^2 \theta\rangle )  . 
\end{equation}
For the alignment process, we have 
\begin{equation} \label{eqap6}
\left. \frac{d}{ds}\langle \cos^2 \theta \rangle \right|_{s_i+0} =
i\langle \varepsilon \lbrack J^2,\cos^2 \theta\rbrack \rangle -4A\varepsilon  \langle\lbrack \cos\theta \sigma_\theta,\cos ^2\theta \rbrack \rangle ,
\end{equation}
which can be rewritten as 
\begin{equation} \label{eqap7}
\left. \frac{d}{ds}\langle \cos^2 \theta \rangle \right|_{s_i+0} =
i\langle \varepsilon \lbrack J^2,\cos^2 \theta\rbrack
\rangle+2A\varepsilon\langle \sin ^2 2\theta\rangle . 
\end{equation} 
From Eq.~(\ref{eqap3}) and from the fact that $\langle
\cos^2\theta\rangle<1$ and $\langle \sin ^2 2\theta \rangle <1$, one
deduces that, in both cases, the slope undergoes a change from zero to
a finite non-zero value when a pulse is applied, for any amplitude $A$
and any maximum of the average of $\cos ^m \theta$. Note that the sign
of the slope and the position of the next local maximum depend on the
sign of $A$.

We address now the case of finite dimensionality. In such a space, the
commutation relations of Eq.~(\ref{eqap2}) cannot be used and we have
to introduce the following standard basis of $su(N)$ \cite{fu}
(related here to the eigenstates of $H_0$),
\begin{eqnarray} \label{appc1}
\left\{
\begin{array}{lll}
x_{mn} &=& e_{mn}-e_{nm}, \\
y_{mn} &=& i(e_{mn}+e_{nm}), \\
h_n &=& i(e_{nn}-e_{n+1,n+1}),
\end{array} \right. 
\end{eqnarray}
where $e_{mn}=|m\rangle\langle n|$, $1\leq n\leq N-1$ and $n< m\leq
N$. $|n\rangle$ is the eigenvector of $H_0$ with eigenvalue $E_n$
[Eq.~(\ref{eq4})]. $H_0$ and the interaction term $H_I$ can be
rewritten with respect to this basis as 
\begin{eqnarray} \label{appc2}
\left\{
\begin{array}{ll}
H_0=\sum_{n=1}^N E_n e_{nn}, \\
H_I=\sum_{n=1}^{N-1} d_n(e_{n,n+1}+e_{n+1,n}),
\end{array} \right. 
\end{eqnarray}
where
\begin{equation}
E_{n+1}=n(n+1)
\end{equation}
and
\begin{equation} \label{appc3}
d_{n+1}=\frac{n+1}{\sqrt{(2n+1)(2n+3)}} .
\end{equation}
Assuming that $\langle [H_0,H_I]\rangle =0$, we now calculate 
\begin{equation} \label{appc3a}
\frac{d}{dt}\langle H_I \rangle =\langle e^{-iAH_I}[H_0,H_I]e^{iAH_I}\rangle ,
\end{equation}
i.e., the derivative of the function $\langle H_I\rangle (t)$ after a
sudden pulse.

A straightforward calculation leads to the identities 
\begin{equation} \label{appc4}
\lbrack H_0,H_I\rbrack  = -\sum_{n=1}^{N-1} \mu_n d_n x_{n,n+1} ,
\end{equation}
where $\mu _n=E_{n+1}-E_n$, and
\begin{eqnarray} \label{appc5}
\frac{1}{2}\lbrack\lbrack H_0,H_I\rbrack ,H_I\rbrack &=& H_I^2
\nonumber \\ & &+\sum_{n=1}^{N-1} (2n+1)d_n^2 e_{n+1,n+1} \nonumber \\ 
& & -\sum_{n=1}^{N} (2n+3)d_n^2 e_{nn} .
\end{eqnarray}
From Eq.~(\ref{appc3}), one can rewrite Eq.~(\ref{appc5}) as
\begin{equation} \label{appc6}
\frac{1}{2}\lbrack\lbrack H_0,H_I\rbrack
,H_I\rbrack=H_I^2-\textbf{1}+\frac{N^2+2N+1}{2N+1}e_{NN} . 
\end{equation}
After expansion according to Campbell-Hausdorf's formula, one gets for the derivative of Eq. (\ref{appc3a}) :
\begin{eqnarray} \label{appc7}
\frac{d}{dt}\langle H_I \rangle & = & 2A(1-\langle
V^2\rangle-\frac{N^2+2N+1}{2N+1} |a_n|^2) \nonumber \\
& &
-i\frac{(N^2+2N+1)N}{(2N+1)^{3/2}\sqrt{2N-1}}A^2(a_N^*a_{N-1}-a_{N-1}^*a_N)
\nonumber \\
& & +O(A^3), 
\end{eqnarray}
where the $a_n$'s are the coefficients of the wave function in the
basis of $|n\rangle$'s. We notice that Eq.~(\ref{appc7}) is very
similar to Eq.~(\ref{eqap5}) except for a few terms. If $A$ is small
enough, the evolution of the average $\langle H_I\rangle$ will depend
on the boundary population $|a_n|^2$.

\section{Analytical estimations of dynamical parameters} \label{appd}
In Sec.~\ref{section5}, the dynamics were investigated through
numerical tests. In this appendix, we give a few analytical
estimations of dynamical parameters such as the time between
pulses. Nevertheless, it is noted that, owing to the complexity of
complete analytical calculations, only rough estimations can be
obtained.

We consider the orientation dynamics and the strategy $S1$. The time
$\Delta t$ between two pulses is assumed to be small enough, which
means that only the last kicks of the sequence will be correctly
described (This point is clearly illustrated in Fig.~\ref{fig4}). We
now analyze the evolution of $\langle\cos\theta\rangle(s)$ between the
times $s_i$ (taken as 0 for simplicity) and $s_f$, a sudden pulse
being applied at $s_i$. We have 
\begin{equation} \label{eqan1}
\langle\cos\theta\rangle(s)=\langle e^{-iA\cos\theta}e^{i\varepsilon J^2 s}\cos\theta e^{-i\varepsilon J^2 s}e^{iA\cos\theta}\rangle ,
\end{equation}
where $s=t/\tau$ is the rescaled time. Expanding the two time
exponentials and neglecting terms of order greater than 2 in $s$,
$\langle\cos\theta\rangle$ can be rewritten in the form
\begin{eqnarray} \label{eqan2}
\langle\cos\theta\rangle(s)&=&\langle\cos\theta\rangle (0) \nonumber \\
& &+i\varepsilon s\langle e^{-iA\cos\theta}\lbrack
J^2,\cos\theta\rbrack e^{iA\cos\theta}\rangle \nonumber \\
& & -\frac{1}{2}\varepsilon ^2 s^2\langle e^{-iA\cos\theta}\lbrack J^2,\lbrack J^2,\cos\theta\rbrack\rbrack e^{iA\cos\theta}\rangle
.\nonumber\\
\end{eqnarray}  
As $\langle \lbrack J^2,\cos\theta\rbrack\rangle (0)=0$, the proof of
Appendix \ref{appb} shows that 
\begin{equation} \label{eqan3}
i\varepsilon s\langle e^{-iA\cos\theta}\lbrack J^2,\cos\theta\rbrack e^{iA\cos\theta}\rangle=2\varepsilon A s (1-\langle \cos ^2\theta\rangle) .
\end{equation}
We next simplify the third term of Eq.~(\ref{eqan2}). Using the
commutation relations \cite{joyeux}
\begin{eqnarray} \label{eqan4}
\left\{ \begin{array}{ll}
\lbrack J^2,\sigma_\theta\rbrack=2\cos\theta J^2, \\
\lbrack J^2,\cos\theta\rbrack=\sigma_\theta+\cos\theta ,
\end{array} \right.
\end{eqnarray}
where $\sigma_\theta=\sin\theta\frac{\partial}{\partial \theta}$, one
arrives at\begin{eqnarray} \label{eqan5}
\langle e^{-iA\cos\theta}\lbrack J^2,\lbrack
J^2,\cos\theta\rbrack\rbrack e^{iA\cos\theta}\rangle &=& 4\langle
\cos\theta J^2\rangle \nonumber \\
&+&8iA\langle \sigma_\theta\cos\theta\rangle \nonumber\\
&+&4A^3\langle
\cos\theta-\cos ^3\theta\rangle \nonumber \\
&+&4iA(1+\langle \cos^2\theta\rangle).\nonumber\\
\end{eqnarray}
Introducing the operators
\begin{eqnarray} \label{eqan6}
\left\{ \begin{array}{ll}
\Re \lbrack \langle \sigma_\theta\cos\theta\rangle \rbrack=-\frac{1}{2}(1+\langle\cos ^2\theta\rangle), \\
\Im \lbrack \langle \sigma_\theta\cos\theta\rangle \rbrack=\frac{1}{2i}\langle \sigma_\theta \cos\theta +\cos\theta \sigma_\theta+2\cos ^2\theta\rangle,
\end{array} \right. 
\end{eqnarray}
where $\Re$ and $\Im$ correspond to the real and the imaginary part of
a complex number, respectively, Eq.~(\ref{eqan2}) becomes
\begin{eqnarray} \label{eqan7}
\langle\cos\theta\rangle (s) &=& \langle\cos\theta\rangle
(0)+2\varepsilon A s(1-\langle\cos
^2\theta\rangle)\nonumber\\
&-&\frac{1}{2}\varepsilon ^2s^2  \lbrack 4\langle \cos\theta J^2\rangle-8A \Im(\langle \sigma_\theta\cos\theta\rangle)\nonumber\\
&+&4A^2\langle \cos\theta-\cos ^3\theta\rangle\rbrack
.
\end{eqnarray}
Equation (\ref{eqan7}) allows us to explicitly determine the time
$s_f$:
\begin{eqnarray} \label{eqan8}
\Delta s&=&s_f-s_i  \nonumber\\
&=&\frac{2A \left(1-\langle\cos
    ^2\theta\rangle\right)}{\varepsilon \lbrack 4\langle \cos\theta
  J^2\rangle-8A \Im(\langle
  \sigma_\theta\cos\theta\rangle)+4A^2\langle \cos\theta-\cos
  ^3\theta\rangle\rbrack} 
.\nonumber\\
\end{eqnarray}
Up to this point, the calculation involves no approximation on the
area of the field $A$, as only the time delays between the pulses are
assumed to be small. In order to highlight different dynamical
behaviors, we now consider the two limits $A\gg 1$ and $A\ll 1$. It
can be shown that
\begin{equation} \label{eqan9}
\Delta s=\frac{1-\langle \cos ^2\theta\rangle}{2\varepsilon A(\langle \cos\theta-\cos^3\theta\rangle )}=\frac{k_{>}}{A} 
\end{equation}
if $A\gg 1$ and
\begin{equation} \label{eqan10}
\Delta s=\frac{A(1-\langle \cos ^2\theta\rangle)}{2\varepsilon \langle\cos\theta J^2\rangle}=k_{<}A 
\end{equation}
if $A\ll 1$. $k_{>}$ and $k_{<}$ are two constants that do not depend
on $A$. They can be estimated by assuming that the wave function is
close enough to the optimal state $|\chi^{(N)}\rangle$. The first
limit corresponds to the strategy suggested in Ref.~\cite{averbuch},
where the area $A$ increases with the number of kicks. From
Eq.~(\ref{eqan9}), we note that the expression of the focusing time
\cite{averbuch,leibscher1,leibscher2} is derived via a quantum
calculation, this time being obtained in Ref.~\cite{leibscher2}
through classical or semi-classical analyses.

\begin{figure}
\includegraphics[width=0.45\textwidth]{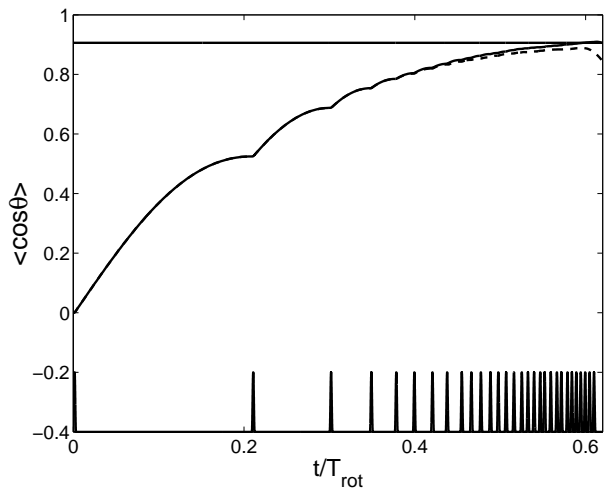}
\caption{\label{fig9} Orientation dynamics during a train of 30
  HCPs. $\varepsilon$ is taken to be 0.01.}
\end{figure}

Due to the moderate intensity of the pulses used in our strategy (for
instance, $A=1$ in Fig.~\ref{fig4}), it can be approximately described
by Eq.~(\ref{eqan10}). We recall that the choice of the strength of
the kicks is crucial for remaining in the finite dimensional subspace
$\mathcal{H}^{(N)}_{m=0}$. This point is clearly illustrated in
Fig.~\ref{fig9}, which gives a view of the orientation dynamics under
the effect of a train of 30 HCPs of area $A=1$.
For the last ten pulses, the arithmetic average of the time delays
between successive kicks is of the order of $\Delta t/T_{rot}\simeq
5.6\times 10^{-3}$. This result can also be derived from the
analytical formula of Eq.~(\ref{eqan10}). Using the eigenstate
$|\chi^{(5)}\rangle$, we obtain $k_<\simeq 5.10^{-3}$ and
\begin{equation} \label{eqan10a}
\frac{\Delta t}{T_{rot}} =\frac{\varepsilon}{\pi}\Delta s=k_<A=5.10^{-3} ,
\end{equation}
as $A=1$. Finally, it is noted that we can also estimate the
difference $\Delta
\langle\cos\theta\rangle=\langle\cos\theta\rangle(s_f)-\langle\cos\theta\rangle(s_i)$
between two successive maxima. Straightforward calculation leads to
\begin{equation} \label{eqan11}
\Delta \langle\cos\theta\rangle =\frac{\left(1-\langle\cos
    ^2\theta\rangle \right)^2A^2}{2\langle \cos\theta J^2\rangle},
\end{equation}
in the hypothesis that $A\ll 1$.


\begin{references}
\bibitem{warrer} W. Warrer, H. Rabitz and M. Dahleh, Science {\bf 259}, 1581 (1993).

\bibitem{seideman} H. Stapelfeldt and T. Seideman,
Rev. Mod. Phys. {\bf 75}, 543 (2003).



\bibitem{atabek} O. Atabek and C. M. Dion and A. Ben Haj Yedder, J. Phys. B {\bf 36}, 4667 (2003).

\bibitem{nielsen} M. A. Nielsen and I. L. Chuang, {\em Quantum computation and quantum information"} (Cambridge Univ. Press, Cambridge, U. K., 2000).

\bibitem{cirac} J. I. Cirac and P. Zoller, Phys. Rev. Lett. {\bf 74}, 4091 (1995).

\bibitem{fonseca} K. M. Fonseca Romera, G. Useche Laverde and F. Torres Ardila, J. Phys. A {\bf 36}, 841 (2003).

\bibitem{dion} C. M. Dion, A. Ben Haj Yedder, E. Canc\`es, A. Keller, C. L. Bris and O. Atabek, Phys. Rev. A {\bf 65}, 063408 (2002).

\bibitem{rabitz} H. Rabitz, R. de Vivie-Riedle, M. Motzkus and K. Kompa, Science {\bf 288}, 824 (2000).

\bibitem{girardeau} M. D. Girardeau, M. Ina, S. Schirmer and T. Gulsrud, Phys. Rev. A {\bf 55}, 3 (1997).

\bibitem{girardeau1} M. D. Girardeau, S. G. Schirmer, J. V. Leahy and R. M. Koch, Phys. Rev. A {\bf 58}, 2684 (1998).
 
\bibitem{rama} V. Ramakrishna, M. V. Salapaka, M. Dahleb, H. Rabitz and A. Peirce, Phys. Rev. A {\bf 51}, 960 (1995).

\bibitem{fu} H. Fu, S. G. Schirmer and A. I. Solomon, J. Phys. A {\bf 34}, 1679 (2001).

\bibitem{sugny} D. Sugny, A. Keller, O. Atabek, D. Daems, C. M. Dion, S. Gu\'erin and H. R. Jauslin, Phys. Rev. A {\bf 69}, 033402 (2004).

\bibitem{vitanov} N. V. Vitanov, T. Halfmann, B. W. Shore and K. Bergmann, Ann. Rev. Phys. Chem. {\bf 52}, 763 (2001).
 
\bibitem{jauslin} S. Gu\'erin and H. R. Jauslin, Adv. Chem. Phys. {\bf 125}, 147 (2003).

 \bibitem{schirmer} S. G. Schirmer, A. D. Greentree, V. Ramakrishna and H. Rabitz, J. Phys. A {\bf 35}, 8315 (2002).
 

\bibitem{zhu} W. Zhu and H. Rabitz, J. Chem. Phys, {\bf 110}, 7142 (1999).

\bibitem{brooks} P. R. Brooks, Science {\bf 193}, 11 (1976).

\bibitem{seideman1} T. Seideman, Phys. Rev. A {\bf 56}, R17 (1997).

\bibitem{seideman2} T. Seideman, J. Chem. Phys. {\bf 111}, 4397 (1999).

\bibitem{aoiz} F. J. Aoiz, Chem. Phys. Lett. {\bf 289}, 132 (1998).

\bibitem{lloyd} S. Lloyd and S. L. Braunstein, Phys. Rev. Lett. {\bf 82}, 1784 (1999).

\bibitem{averbuch} I. S. Averbukh and R. Arvieu, Phys. Rev. Lett. {\bf 87}, 163601 (2001).

\bibitem{vrakking} M. J. J. Vrakking and S. Solte, Chem. Phys. Lett. {\bf 271}, 209 (1997).

\bibitem{guerin} S. Gu\'erin , L. P. Yatsenko, H. R. Jauslin, O. Faucher and B. Lavorel, Phys. Rev. Lett. {\bf 88}, 233601 (2002).

\bibitem{friedrich} B. Friedrich and D. Herschbach, Phys. Rev. Lett. {\bf 74}, 4623 (1995).

\bibitem{sangouard} N. Sangouard, S. Gu\'erin, M. Amniat-Talab and H. R. Jauslin, Phys. Rev. Lett. {\bf 93}, 223602 (2004).

\bibitem{dion2} C. M. Dion, A. Keller and O. Atabek, Eur. Phys. J. D {\bf 14}, 249 (2001).

\bibitem{henriksen} N. E. Henriksen, Chem. Phys. Lett. {\bf 312}, 196 (1999).
 
\bibitem{keller} A. Keller, C. M. Dion and O. Atabek, Phys. Rev. A {\bf 61}, 023409 (2000).

 
\bibitem{dion1} C. M. Dion, A. D. Bandrauk, O. Atabek, A. Keller, H. Umeda and Y. Fujimura, Chem. Phys. Lett. {\bf 302}, 215 (1999).

\bibitem{machholm} M. Machholm and N. E. Henriksen, Phys. Rev. Lett. {\bf 87}, 193001 (2001).

\bibitem{sugny2} D. Sugny, A. Keller, O. Atabek, D. Daems, S. Gu\'erin and H. R. Jauslin, Phys. Rev. A {\bf 69}, 043407 (2004).


\bibitem{daems} D. Daems, A. Keller, S. Gu\'erin, H. R. Jauslin and O. Atabek, Phys. Rev. A {\bf 67}, 052505 (2003).

\bibitem{machholm1} M. Machholm, J. Chem. Phys. {\bf 115}, 10724 (2001).

\bibitem{ortigoso} J. Ortigoso, M. Rodriguez, M. Gupta and B. Friedrich, J. Chem. Phys. {\bf 110}, 3870 (1999).

\bibitem{benhajyedder} A. Ben Haj Yedder, A. Auger, C. M. Dion, 
                  E. Canc\`es, A. Keller, C. Le Bris and
                  O. Atabek, Phys. Rev. A {\bf 66}, 063401 (2002).

\bibitem{shlomo} Shlomo E. Sklarz, D. J. Tannor and N. Khaneja, Phys. Rev. A {\bf 69}, 053408 (2004).

\bibitem{tannor} D. J. Tannor and A. Bartana, J. Phys. Chem. A {\bf 103}, 10359 (1999).

\bibitem{beige} A. Beige, D. Braun, B. Tregenna and P. L. Knight, Phys. Rev. Lett. {\bf 85}, 1762 (2000).
\bibitem{lidar} D. A. Lidar, D. Bacon, J. Kempe and K. B. Whaley, Phys. Rev. A  {\bf 63}, 022307 (2001).


\bibitem{merzbacher} E. Merzbacher, {\em Quantum Mechanics} (Wiley, New York, 1970).

\bibitem{schirmer2} S. G. Schirmer, H. fu and A. I. Solomon, Phys. Rev. A {\bf 63}, 063410 (2001).

\bibitem{joyeux} M. Joyeux and D. Sugny, Can. J. Phys. {\bf 80}, 1459 (2002).


\bibitem{leibscher1} M. Leibscher, I. Sh. Averbukh and H. Rabitz, Phys. Rev. Lett. {\bf 90}, 213001 (2003).

\bibitem{leibscher2} M. Leibscher, I. Sh. Averbukh, P. Rozmej and R. Arvieu, Phys. Rev. A {\bf 69}, 032102 (2004).

 
 


\end{references}
\end{document}